\documentclass{article}
\usepackage{arxiv}

\usepackage[utf8]{inputenc} 
\usepackage[T1]{fontenc}    
\usepackage{algorithmic}
\usepackage{graphicx}
\usepackage{subcaption}
\usepackage{comment}
\usepackage{ulem}
\usepackage{url}            
\usepackage[hyphenbreaks]{breakurl}
\usepackage{stfloats}
\usepackage{fixltx2e}
\usepackage{textcomp}
\usepackage[table]{xcolor}
\usepackage{colortbl}
\usepackage{siunitx}
\usepackage{glossaries}
\usepackage{booktabs}       
\usepackage{amsfonts}       
\usepackage{nicefrac}       
\usepackage{microtype}      
\usepackage{lipsum}		
\usepackage[numbers]{natbib}
\usepackage{doi}

\newacronym{asd}{ASD}{anomalous sound detection}
\newacronym{xai}{XAI}{Explainable AI}

\title{A Framework for Evaluating Faithfulness in Explainable AI for Machine Anomalous Sound Detection Using Frequency-Band Perturbation}

\author{
Alexander Buck$^{1}$,
Georgina Cosma$^{1}$,
Iain Phillips$^{1,2}$,
Paul Conway$^{3}$,
Patrick Baker$^{2,4}$
}

\date{}

\renewcommand{\headeright}{}
\renewcommand{\undertitle}{\textit{arXiv} Preprint}
\renewcommand{\shorttitle}{\textit{arXiv} Preprint}

\hypersetup{
pdftitle={A template for the arxiv style},
pdfsubject={q-bio.NC, q-bio.QM},
pdfauthor={David S.~Hippocampus, Elias D.~Striatum},
pdfkeywords={First keyword, Second keyword, More},
}

\begin{document}

\maketitle 
\vspace{-3.5em}   

\begin{center}
$^{1}$Computer Science, School of Science, Loughborough University, UK\\
$^{2}$Royal Air Force Rapid Capabilities Office, UK\\
$^{3}$School of Mechanical, Electrical and Manufacturing Engineering, Loughborough University, UK\\
$^{4}$Defence Science and Technology Laboratory, UK
\end{center}
\vspace{3.5em}   

\begin{abstract}

Explainable AI (XAI) is commonly applied to anomalous sound detection (ASD) models to identify which time–frequency regions of an audio signal contribute to an anomaly decision. However, most audio explanations rely on qualitative inspection of saliency maps, leaving open the question of whether these attributions accurately reflect the spectral cues the model uses. In this work, we introduce a new quantitative framework for evaluating XAI faithfulness in machine-sound analysis by directly linking attribution relevance to model behaviour through systematic frequency-band removal. This approach provides an objective measure of whether an XAI method for machine ASD correctly identifies frequency regions that influence an ASD model’s predictions. By using four widely adopted methods, namely Integrated Gradients, Occlusion, Grad-CAM and SmoothGrad, we show that XAI techniques differ in reliability, with Occlusion demonstrating the strongest alignment with true model sensitivity and gradient-based methods often failing to accurately capture spectral dependencies. The proposed framework offers a reproducible way to benchmark audio explanations and enables more trustworthy interpretation of spectrogram-based ASD systems.

\end{abstract}


\section{Introduction}

In \gls{asd}, model predictions are often based on subtle spectral patterns, making it essential to understand which parts of an audio signal drive an anomaly decision. \gls{xai} methods are therefore used to attribute model decisions to specific time–frequency components of the input. Much of the existing work in audio \gls{xai} relies on visual inspection of attribution maps over spectrograms, assuming that visually plausible heatmaps meaningfully reflect the cues the model uses \cite{11205342}. Yet despite the rapid growth of interest in this area, the field still lacks a clear understanding of whether these explanations faithfully correspond to the spectral regions that drive a model’s decisions. This problem is compounded by the fact that various \gls{xai} methods routinely disagree with one another, while offering neither accuracy measure nor objective criterion for determining which explanation should be trusted \cite{doi:10.34133/icomputing.0074}.

These limitations highlight a significant gap: there is currently no quantitative framework for assessing the faithfulness of \gls{xai} methods in audio, and no way to verify whether an attribution map aligns with the true acoustic evidence used by an \gls{asd} model. Prior work in other domains has introduced perturbation-based \cite{10.1007/978-3-031-20319-0_30} or behavioural validation techniques \cite{Lee2025-te}, but such approaches have not yet been adapted to the machine-sound setting, where frequency-dependent behaviour is central to model interpretation.

In this work, we address this gap by conducting the first systematic, quantitative investigation of \gls{xai} faithfulness for machine-sound anomaly detection. We compare four widely used attribution methods on a spectrogram-based \gls{asd} model, analysing how their relevance maps differ and where their explanations diverge. To identify which frequency bands the model relies on for its anomaly predictions, we perform a systematic frequency-band removal analysis that measures changes in the anomaly score when specific spectral regions are suppressed. The resulting change in the model’s output quantifies the model's sensitivity to each band, providing a functional ground truth for the band’s importance with respect to the model’s anomaly score. We then introduce a faithfulness evaluation framework that directly links \gls{xai} relevance to model sensitivity: for each frequency band, we correlate the mean \gls{xai}-assigned relevance with the prediction change induced by band removal. This allows us to determine, for the first time, whether an \gls{xai} method faithfully captures the spectral features the model uses.

By combining attribution analysis with perturbation-based validation, the proposed framework provides a clear and systematic way to evaluate audio \gls{xai} methods. The framework identifies techniques that reliably reflect model behaviour and those that produce misleading explanations, offering new guidance for the development and evaluation of interpretable \gls{asd} systems.

\section{Background and Related Work}

\subsection{Anomalous Sound Detection (ASD)}

\gls{asd} aims to identify deviations from normal machine operation using primarily normal-only training data, making it fundamentally a semi or self-supervised problem. \gls{asd} pipelines typically convert raw audio into time–frequency representations such as log-Mel \cite{10942268} or magnitude spectrograms \cite{9632762}, on which models learn the distribution of normal machine behaviour.

Neural architectures for \gls{asd} often rely on convolutional encoders \cite{zhou2025machine} or autoencoder-style embeddings \cite{10557607} to characterise normal acoustic patterns. Recent work has highlighted the advantages of self-supervised representation learning, where models learn discriminative embeddings without anomalous labels and achieve strong robustness under domain shift \cite{10447156, 11150719}. These approaches address core \gls{asd} challenges including variable acoustic environments, machine-to-machine variation, and frequency-dependent characteristics of mechanical systems \cite{Harada2021}.

Because \gls{asd} models can be employed in safety-critical industrial monitoring, interpretability plays a central role. False positives can trigger unnecessary inspections, while false negatives may allow mechanical faults to progress undetected. Interpretable systems help practitioners verify model behaviour, understand which spectral cues drive anomaly scores, and assess whether the model relies on meaningful physical signatures rather than noise or artefacts, all of which are common problems in \gls{asd} \cite{10289804}.

\subsection{XAI for Audio}

\gls{xai} for audio is considerably less developed than in vision or natural language processing, where attribution methods have been extensively benchmarked and validated. In audio classification and machine-condition monitoring, explanations typically operate on spectrogram representations, allowing the use of image-based attribution methods such as Integrated gradients \cite{pmlr-v70-sundararajan17a}, Grad-CAM \cite{8237336}, Occlusion \cite{Fong_2017_ICCV}, and SmoothGrad \cite{smilkov2017smoothgradremovingnoiseadding}, which have been adapted from computer vision pipelines to the time–frequency domain \cite{Selvaraju_2017_ICCV}.

Despite their widespread use, existing audio \gls{xai} studies predominantly rely on qualitative inspection of saliency maps over spectrograms. Prior work has repeatedly noted that visual attribution maps can vary substantially between methods, often highlighting different or inconsistent regions of the input even when applied to the same model \cite{govindu2023deepfake}. This disagreement between methods makes it difficult to determine which explanations accurately reflect model behaviour, particularly in safety-critical acoustic applications such as machine sound diagnostics.

The main limitation of \gls{xai} in machine audio is the absence of quantitative evaluation frameworks for assessing whether an attribution map corresponds to the spectral cues a model uses. Heatmaps may appear intuitively plausible, yet there is no guarantee that these highlighted areas meaningfully influence the model’s predictions. Recent work in related areas such as speech forensics has emphasised this problem, noting that visually convincing saliency does not necessarily imply explanation correctness without behavioural validation \cite{10289804}. This gap motivates the need for objective, model-centric metrics for \gls{xai} in audio.

\subsection{Faithfulness and Evaluation of XAI Methods}

A central principle in \gls{xai} is faithfulness, which refers to how accurately an attribution method reflects the true decision process of a model. Faithfulness is distinct from plausibility, which concerns whether an explanation looks intuitive to a human observer \cite{wojciechowski-etal-2024-faithful}. An attribution map may appear reasonable yet still fail to represent the features the model relies on. This distinction has been widely emphasised across \gls{xai} research, particularly in cases where explanations must inform safety-critical decisions \cite{afolabi2025faithful}.

To assess faithfulness, several evaluation strategies have been developed in other domains. A common approach is perturbation-based validation, where parts of the input are systematically removed or masked to measure how the model’s output changes \cite{AMORIM2023103225}. Related techniques include sensitivity analysis \cite{LAMPROU2024108238}, which tests whether small or semantically meaningful input changes yield proportional attribution responses, and occlusion tests \cite{10912611}, where local input regions are replaced or corrupted to estimate their functional contribution to model predictions. Researchers have widely applied these methods in computer vision, NLP, and speech processing to quantify the extent to which an explanation aligns with model behaviour \cite{MAMALAKIS202570}. 

However, these faithfulness evaluation strategies have not yet been adapted to machine sound anomaly detection, where spectral structure and frequency-dependent behaviour play a central role. In particular, while perturbation tests have been used in speech tasks, no prior work has applied frequency-band–level perturbation to validate whether spectrogram-based \gls{xai} methods correctly identify the spectral regions driving \gls{asd} model decisions. This gap motivates the need for quantitative, frequency-aware evaluation frameworks tailored specifically to machine-sound analysis.

\subsection{Frequency-Based Analysis for Audio Models}

A small number of recent studies have explored frequency-band masking as a way to analyse model behaviour in audio tasks. For example, Salvi et al. \cite{10289804} applied band-level perturbations to assess which spectral regions were most informative for synthetic speech detection, demonstrating that selective frequency removal can reveal model biases and sensitivity patterns. Similar techniques have appeared in speech forensics and anti-spoofing research, where frequency-focused perturbations are used to probe robustness under domain shifts \cite{electronics11142183}.

However, these works are not \gls{xai} faithfulness evaluations. They do not compare attribution maps with the model’s behavioural response to perturbations, nor do they assess whether explanation methods accurately reflect the spectral cues a model relies upon. Prior studies have focused primarily on speech and synthetic audio, where spectral characteristics differ substantially from mechanical sound environments. None address the machine-sound anomaly detection setting, where frequency-dependent behaviour is central to interpreting model decisions.

As a result, there remains no existing framework that quantitatively evaluates the faithfulness of \gls{xai} methods in machine-sound analysis using frequency-band perturbations. This gap motivates the contribution of the present work, which integrates attribution methods with systematic spectral perturbations to assess whether explanations align with a model’s true dependence on specific frequency regions.

\section{Proposed Framework for Testing the Faithfulness of XAI Methods in Machine ASD}

\begin{figure*}[htbp]
    \centering
    \includegraphics[width=1\columnwidth]{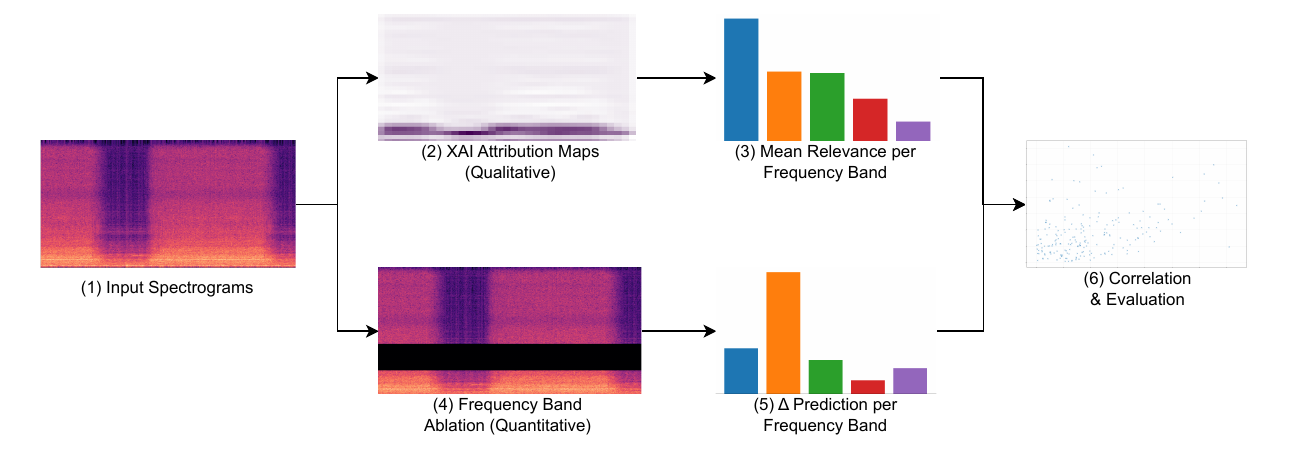}
    \caption{Framework for evaluating \gls{xai} methods on a machine audio anomaly detection task.}
    \label{fig:proposal}
\end{figure*}

In this work, we propose a new framework for the quantitative evaluation of \gls{xai} methods in \gls{asd}. 

Figure \ref{fig:proposal} shows the proposed framework. Existing studies in audio \gls{xai} typically rely on qualitative inspection of attribution maps over spectrograms, leaving open the question of how accurately these attribution techniques reflect the true spectral regions that influence a model’s predictions. Our method addresses this gap by introducing a repeatable procedure that directly links attribution relevance to the model’s actual sensitivity to different frequency components. This enables a more rigorous assessment of \gls{xai} methods in machine-sound diagnostics and offers insights into the decision-making behaviour of \gls{asd} models.

The numbers in parentheses (e.g., 1) refer to the corresponding steps shown in the framework diagram. The input to the framework is a sound spectrogram, generated from a raw sound waveform (1). Then, four commonly used attribution methods are applied to an existing \gls{asd} model trained on spectrogram representations (2). These \gls{xai} methods produce time–frequency relevance maps that highlight the regions of the input the model considers important. Each attribution map is then converted into band-level mean relevance scores (3). Each attribution map is normalised so that all relevance values are non-negative and sum to one, then the mean relevance assigned to each of the same frequency bands used in the perturbation analysis is computed. 

While such heatmaps provide valuable qualitative insights, they do not indicate whether the highlighted regions correspond to the frequency bands the model truly relies upon. To construct a quantitative baseline for comparison, a systematic frequency-band perturbation analysis is performed. In this step, the original input (1) is used. Predefined frequency intervals are replaced with zeros (e.g., \SIrange{0}{1600}{\hertz}, \SIrange{1600}{3200}{\hertz}, etc.) and the model is retrained and retested on these altered inputs (4). For each ablated band, the change in prediction relative to the model trained and tested on the full spectrogram is computed. This produces a $\Delta$-prediction curve that captures the sensitivity of the model to each frequency region and serves as a reference measure of true spectral importance (5).

The agreement between attribution-based relevance and $\Delta$-prediction sensitivity is measured using Spearman’s rank correlation coefficient. This correlation quantifies the extent to which each \gls{xai} method correctly identifies the frequency bands that the model depends on. High correlation indicates that the \gls{xai} method faithfully explains model behaviour, while low correlation reveals mismatches between perceived and actual spectral importance.

\section{Experimental Methodology}

This section describes the experimental setup and procedures used throughout the paper. Sections~\ref{sec:dataset}--\ref{sec:spectrogram} detail the dataset, model architecture, and spectrogram extraction process, that are shared across all experiments. Sections~\ref{sec:methodology 1}--\ref{sec:methodology 3} then describe three experiments, each designed to address a specific aspect of XAI evaluation. Experiment 1 and Experiment 2 establish controlled perturbation-based analyses of the model’s behaviour, while Experiment 3 builds upon these results to quantitatively evaluate the faithfulness of the considered \gls{xai} methods. Specifically, Section~\ref{sec:methodology 1} corresponds to Experiment 1 (reported in Section~\ref{sec:experiment 1}), Section~\ref{sec:methodology 2} to Experiment 2 (reported in Section~\ref{sec:experiment 2}), and Section~\ref{sec:methodology 3} to Experiment 3 (reported in Section~\ref{sec:experiment 3}).

\subsection{Dataset}
\label{sec:dataset}

All experiments in this study are carried out using the DCASE2023 Task 2 \gls{asd} dataset \cite{dohi2023description}, which provides noisy recordings of machine operating sounds intended for semi-supervised machine-condition monitoring. The audio originates from the ToyAdmos2 dataset \cite{Harada2021} and MIMII-DG \cite{Dohi2022}, both of which contain recordings collected under varied acoustic conditions and machine configurations. Although the original benchmark is structured around source and target domains to evaluate domain generalisation performance, this distinction is not retained in the present work. Instead, all available normal recordings are pooled into a single training set, as the objective here is not few-shot or first-shot anomaly detection but rather to investigate model behaviour with different \gls{xai} methods applied.

The dataset is divided into a development set and an evaluation set. Each split contains a training subset consisting entirely of normal sounds and a test subset that includes both normal and anomalous samples. The DCASE2023 Task 2 dataset has 14 distinct machine types, with each type represented by a single machine ID. Furthermore, the machine types assigned to the development and evaluation portions are different, which is intended to test generalisation across previously unseen machine categories. This separation of machine types, combined with the limited diversity within each category, makes learning discriminative representations from classification tasks very challenging. As a result, the dataset naturally motivates the use of representation-learning or self-supervised approaches when developing \gls{asd} systems. The characteristics of the dataset can be seen in Table~\ref{dataset characteristics}.

\begin{table}[!h]
    \centering
    \begin{tabular}{lr}
            \toprule
            Dataset	&	DCASE 2023 Task 2 \cite{dohi2023description} 	\\
            \midrule
            Machine Types	&	14	 \\
            Train Size (per machine)     &   1000 \\
            Test Size (per machine)      &   200  \\
            No. Channels	&	1	 \\
            Sample Length	&	6–18 seconds	\\
            Sample Rate     &   \SI{16}{\kilo\hertz} \\
         \bottomrule
    \end{tabular}
    \caption{Dataset characteristics}
    \label{dataset characteristics}
\end{table}

\subsection{ASD Model Architecture}
\label{sec:model}

In this work, the publicly documented model proposed by Wilkinghoff et al. \cite{10447156} in Self-Supervised Learning for Anomalous Sound Detection is used. This model does not include any \gls{xai} functionality. Model architecture, training procedure, and anomaly-scoring pipeline are used exactly as described in the original publication and accompanying implementation resources. We select the model introduced by Wilkinghoff et al., as its performance and generalisation capabilities have already been extensively evaluated against competing approaches on the DCASE 2023 Task 2 dataset. By relying on a previously benchmarked and domain-validated architecture, we ensure that our analysis of \gls{xai} faithfulness is conducted in a realistic and practically relevant setting, while avoiding the need for additional model-level comparisons. As the model itself is not the focus of our contribution, we provide only a brief summary below; full architectural details can be found in \cite{10447156}.

The model follows a self-supervised feature-learning paradigm based on the Feature Exchange (FeatEx) approach. A convolutional encoder processes magnitude spectrograms and produces an embedding vector, which is trained using a self-supervised contrastive-style loss augmented with the FeatEx approach. During inference, anomaly scores are computed from embedding-space distances between input samples and the distribution of normal training embeddings. We match all hyperparameters, augmentation settings, and training procedures from the original work 
\cite{10447156} unless stated otherwise.

\subsection{Spectrogram extraction}
\label{sec:spectrogram}

The \gls{xai} methods in this study analyse a neural network that operates on magnitude spectrogram representations. To ensure reproducibility and enable clear interpretation of results, we detail the spectrogram computation pipeline used throughout model training and evaluation.
Given a discrete-time audio signal $x[n]$, the time-frequency representation is computed via the Short-Time Fourier Transform (STFT):
\begin{equation}
X[m,k] = \sum_{n=0}^{N-1} x[n]w[n - mH]e^{-j2\pi kn/N}
\end{equation}
where $w[n]$ is the analysis window (Hamming window in our implementation), $H$ is the hop size in samples, $N$ specifies the FFT size, $m$ indexes time frames, and $k$ indexes frequency bins. To compute the magnitude spectrogram, the absolute value is computed:
\begin{equation}
S[m,k] = |X[m,k]|
\end{equation}

\begin{figure}[htbp]
    \centering
    \includegraphics[width=1\columnwidth]{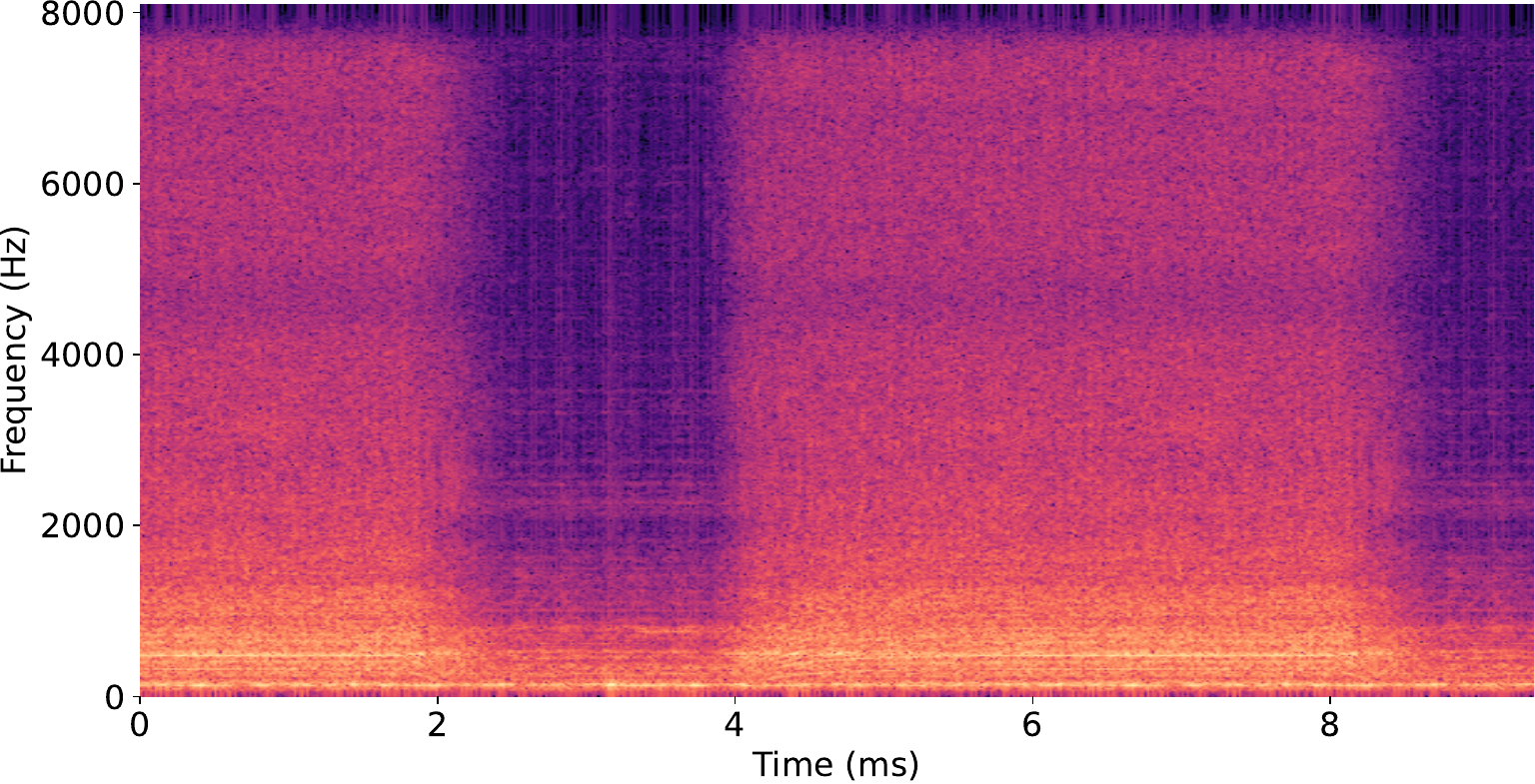}
    \caption{Spectrogram representation of an audio signal, with brighter colours representing higher magnitude sounds and darker colours representing lower magnitude sounds.}
    \label{fig:spectrogram}
\end{figure}

These magnitude spectrograms serve as input features for our neural network (Section 5.2). Figure~\ref{fig:spectrogram} shows an example of a magnitude spectrogram, showing time on the horizontal axis and frequency on the vertical axis. Brighter regions indicate higher energy, while darker regions correspond to lower-intensity components.

\subsection{XAI methods evaluated}
\label{sec:methodology 1}

Standard image-based \gls{xai} methods are applied to spectrogram representations of audio. Although originally developed for vision tasks, these attribution techniques are now widely used in audio because spectrograms provide a natural 2-D structure for analysing time–frequency patterns in machine sound. Their popularity is driven by two factors: spectrograms are already the dominant input representation in \gls{asd} systems, and image-based \gls{xai} methods are mature, well-supported, and straightforward to integrate into existing pipelines. In industrial machine-sound analysis, interpretability is primarily aimed at technical users, making spectrogram-based visual explanations an appropriate and practical choice. Table~\ref{tab:xai_features} shows the key features of the four \gls{xai} methods that are used.

\begin{table*}[t]
    \centering
    \caption{Comparison of \gls{xai} methods for explaining sound spectrograms}
    \label{tab:xai_features}
    \begin{tabular}{l *{4}{p{1.2in}}}
        \toprule
        \textbf{Feature} & \textbf{Grad-CAM} & \textbf{Integrated Gradients} & \textbf{Occlusion} & \textbf{SmoothGrad} \\
        \midrule
        \textbf{Explanation Type} & Local & Local & Local & Local \\
        \textbf{Model Dependency} & Yes (CNN-specific) & Yes & No & Yes \\
        \textbf{Computational Cost} & Low & High & Very High & High \\
        \textbf{Perturbation-Based} & No & No & Yes & Yes \\
        \textbf{Gradient-Based} & Yes & Yes & No & Yes \\
        \textbf{Resolution} & Low (layer-dependent) & High (pixel-level) & Medium (window size) & High (pixel-level) \\
        \textbf{Common Use Cases} & CNN visualisation, Class activation & Feature attribution, Sensitivity analysis & Feature importance, Robustness testing & Noise reduction, Saliency refinement \\
        \textbf{Key Advantages} & Class-specific, No model retraining & Complete, Theoretical guarantees & Model-agnostic, Intuitive & Reduces noise, Sharpens attributions \\
        \textbf{Limitations} & Limited to CNNs, Coarse localisation & Baseline choice sensitive, Computationally expensive & Computationally intensive, Window size sensitive & Adds hyperparameters, Computationally expensive \\
        \bottomrule
    \end{tabular}
\end{table*}
\textbf{Grad-CAM} generates class-specific relevance maps by weighting the spatial activation maps of a chosen convolutional layer with the gradients of the output class score. Because Grad-CAM operates directly on feature maps rather than raw inputs, the resulting heatmaps highlight which high-level spectro-temporal patterns most strongly contribute to the model’s decision. This makes Grad-CAM well suited for visualising CNN attention over sound spectrograms, although its reliance on convolutional layers limits its applicability to CNN-based architectures. Moreover, Grad-CAM typically produces coarse localisation since it inherits the spatial resolution of the feature maps rather than the input spectrogram.

\textbf{Integrated Gradients} computes feature attributions by accumulating gradients along a straight-line path between a baseline input and the actual spectrogram. Following common practice, a zero-valued baseline is used to represent an absence of acoustic energy, and the attributions for each input feature are computed by integrating the gradients of the model's output with respect to the input along a path from the baseline to the actual input. Compared to raw gradients, this yields attribution maps that satisfy completeness and are less sensitive to gradient noise. However, the Integrated Gradients method is computationally more expensive due to the need for multiple gradient evaluations, and results depend on the chosen baseline, which may introduce ambiguity for certain spectrogram structures.

\textbf{Occlusion} estimates feature importance by systematically masking out small patches of the input spectrogram and observing changes in the model’s output. By measuring how prediction confidence degrades when specific time–frequency regions are removed, the method provides intuitive, model-agnostic attributions. In our setup, Occlusion was applied using sliding windows over the spectrogram, directly revealing which localised acoustic events or spectral bands the network relies on. Despite its interpretability, the method is computationally intensive and can be sensitive to the choice of window size, which must balance spatial resolution and computational cost.

\textbf{SmoothGrad} refines gradient-based explanations by averaging saliency maps computed from multiple noisy variants of the input spectrogram. Gaussian noise with zero mean and small variance is added to create perturbed spectrograms, generating gradients for each and averaging the resulting attribution maps. This procedure reduces high-frequency noise present in single-gradient explanations and yields sharper, more stable saliency patterns. While SmoothGrad is effective at highlighting consistent structures, such as harmonic ridges or transient onsets, it introduces additional hyperparameters, and increases computational cost due to repeated gradient evaluations. 

\subsection{Frequency-band importance analysis}
\label{sec:methodology 2}

In this section, the contribution of different frequency bands to machine sound anomaly detection performance on the DCASE 2023 Task 2 dataset is investigated. The objective of this approach is to evaluate the sensitivity of the model to specific frequency ranges and to identify which parts of the spectrum are most informative for detecting anomalies. This type of frequency-band analysis has been shown to provide insights into model robustness and interpretability, as it highlights the frequency regions most relevant for classification decisions \cite{10289804}.

To perform this analysis, the same model architecture is used as in previous sections as a baseline, using both spectrum and spectrogram inputs. A systematic frequency-masking procedure is applied, following the methodology of Salvi et al. \cite{10289804}. For each machine type and condition, modified versions of the input signals are constructed by excluding successive frequency bands of width \SI{1600}{\hertz}: \SIrange{0}{1600}{\hertz}, \SIrange{1600}{3200}{\hertz}, \SIrange{3200}{4800}{\hertz}, \SIrange{4800}{6400}{\hertz}, and \SIrange{6400}{8000}{\hertz}. Exclusion is then achieved by setting the normalised magnitude components within the selected frequency bands to zeros within both the spectral and spectrogram representations, thereby removing the information from those bands while leaving the remaining spectral content unchanged. This procedure allows us to quantify the degree to which model performance depends on each frequency band.

\subsection{Faithfulness evaluation of XAI methods}
\label{sec:methodology 3}

We introduce a method to quantitatively assess the faithfulness of the \gls{xai} methods used in previous sections. These are: Integrated Gradients, Occlusion, Grad-CAM, and SmoothGrad. This will be done by testing whether the relevance attributed to specific frequency bands corresponds to the model’s actual dependence on those bands for prediction. A strong correlation between the mean relevance scores of frequency regions and the resulting change in the model’s output when those regions are removed would demonstrate that the \gls{xai} method provides faithful explanations consistent with the model’s internal reasoning.

\subsubsection{Mean Relevance per Frequency Band}

For each audio sample, a spectrogram was divided into five frequency bands: \SIrange{0}{1600}{\hertz}, \SIrange{1600}{3200}{\hertz}, \SIrange{3200}{4800}{\hertz}, \SIrange{4800}{6400}{\hertz}, and \SIrange{6400}{8000}{\hertz}.
Given a relevance map $R(t, f)$ generated by an \gls{xai} method, where $t$ indexes time frames and $f$ indexes frequency bins, the mean relevance within a specific band [$f_{\mathrm{low}}$,$f_{\mathrm{high}}$] is computed as:
\[
\bar{R}_{[f_{\mathrm{low}}, f_{\mathrm{high}}]} = \frac{1}{T \cdot (f_{\mathrm{high}} - f_{\mathrm{low}} + 1)} \sum_{t=1}^{T} \sum_{f = f_{\mathrm{low}}}^{f_{\mathrm{high}}} R(t, f)
\]

where T denotes the number of time frames in the spectrogram. This measure provides an aggregated relevance score that captures the average contribution of each frequency range to the model’s prediction.

\subsubsection{Change in Model Prediction ($\Delta$Pred)}

To quantify the model’s sensitivity to each frequency band, a band-masking approach is used. For every sample, a new spectrogram is generated with the normalised magnitude components of one frequency band set to zero while the rest remained unchanged. The difference between the model’s output for the original and modified spectrograms ($\Delta$Pred) is then computed as:
\[
\Delta \mathrm{Pred} = |\mathrm{Pred}_{\mathrm{original}} - \mathrm{Pred}_{\mathrm{sample}}|
\]
A large $\Delta$Pred implies that the removed frequency region had a strong effect on the model’s decision.

\subsubsection{Correlation Analysis}

For each machine type and frequency band, the Spearman rank correlation coefficient ($\rho_s$) is calculated between the vectors of mean relevance scores $\bar{R}$ assigned to each of the five predefined frequency bands and the corresponding $\Delta \mathrm{Pred}$ observed when each band is masked. Each paired observation represents a single test sample, where the mean relevance within each frequency band is compared with the change in model prediction resulting from masking the corresponding band. The formula for Spearman rank correlation coefficient is:
\[
\rho_s = 1 - \frac{6 \sum_{i=1}^{n} d_i^2}{n(n^2 - 1)}
\]
where $d_{\mathrm{i}}$ represents the difference between the ranks of the i-th sample according to $\bar{R}$ and $\Delta \mathrm{Pred}$, and n is the number of paired observations. Spearman correlation has been chosen because it evaluates whether frequency bands ranked as more relevant by an \gls{xai} method also produce larger changes in the model’s prediction when masked. This metric does so without assuming a linear relationship between relevance magnitude and prediction change, and while remaining robust to outlier samples. Significance levels are computed using two-tailed p-values to evaluate whether the observed correlations are statistically different from zero. 

Because correlation coefficients are not additive in their raw form, the Fisher z-transformation is applied when computing the overall average correlation across the 200 test samples. Each correlation value $\rho_s$ is first transformed using:
\[
z = \frac{1}{2}\ln\left(\frac{1 + \rho_s}{1 - \rho_s}\right),
\]
which stabilises the variance and yields approximately normally distributed values. The overall mean correlation is then obtained by averaging the transformed scores and converting back to the correlation scale using the inverse transform:
\[
\bar{\rho}_s = \frac{e^{2\bar{z}} - 1}{e^{2\bar{z}} + 1},
\]
where $\bar{z}$ denotes the mean of the Fisher-transformed correlations, and $\bar{\rho_s}$ is our final faithfulness metric. This procedure provides a statistically well-behaved estimate of the aggregate correlation for each \gls{xai} method.

\section{Experiment 1: Qualitative Comparison of XAI Methods}
\label{sec:experiment 1}

\begin{figure*} 
\centering
    \caption{Visualisation of \gls{xai} methods on audio spectrograms. Each column represents a different sample from the test set. The first row shows the original audio spectrograms, and the subsequent rows represent the relevance attributions of the \gls{xai} methods in the following order: Integrated Gradients (second row), Occlusion (third row), Grad-CAM (fourth row), and SmoothGrad (fifth row). Regions of interest (highlighted in purple) were generated using the \gls{xai} methods described in this study.}
\label{fig:five_images}

\newlength{\mywidth}
\setlength{\mywidth}{0.24\textwidth}

\vspace{4pt}
\includegraphics[width=\mywidth]{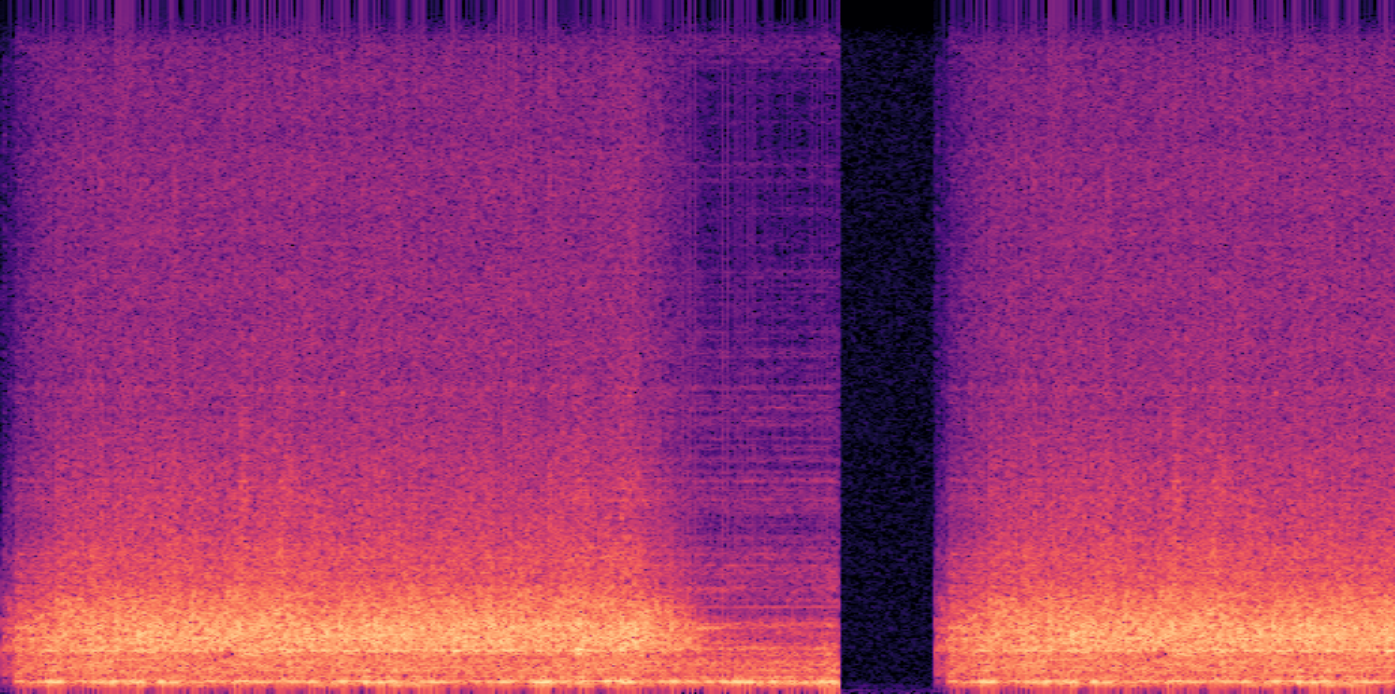}\hfill
\includegraphics[width=\mywidth]{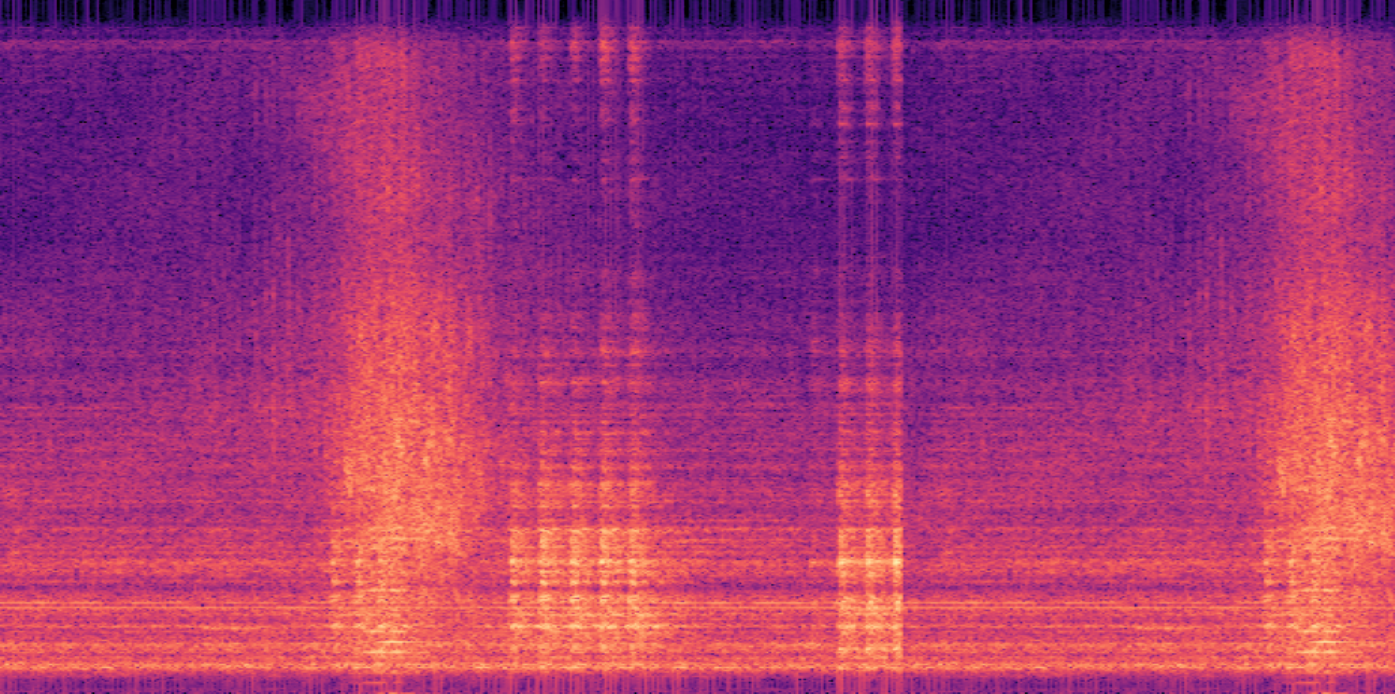}\hfill
\includegraphics[width=\mywidth]{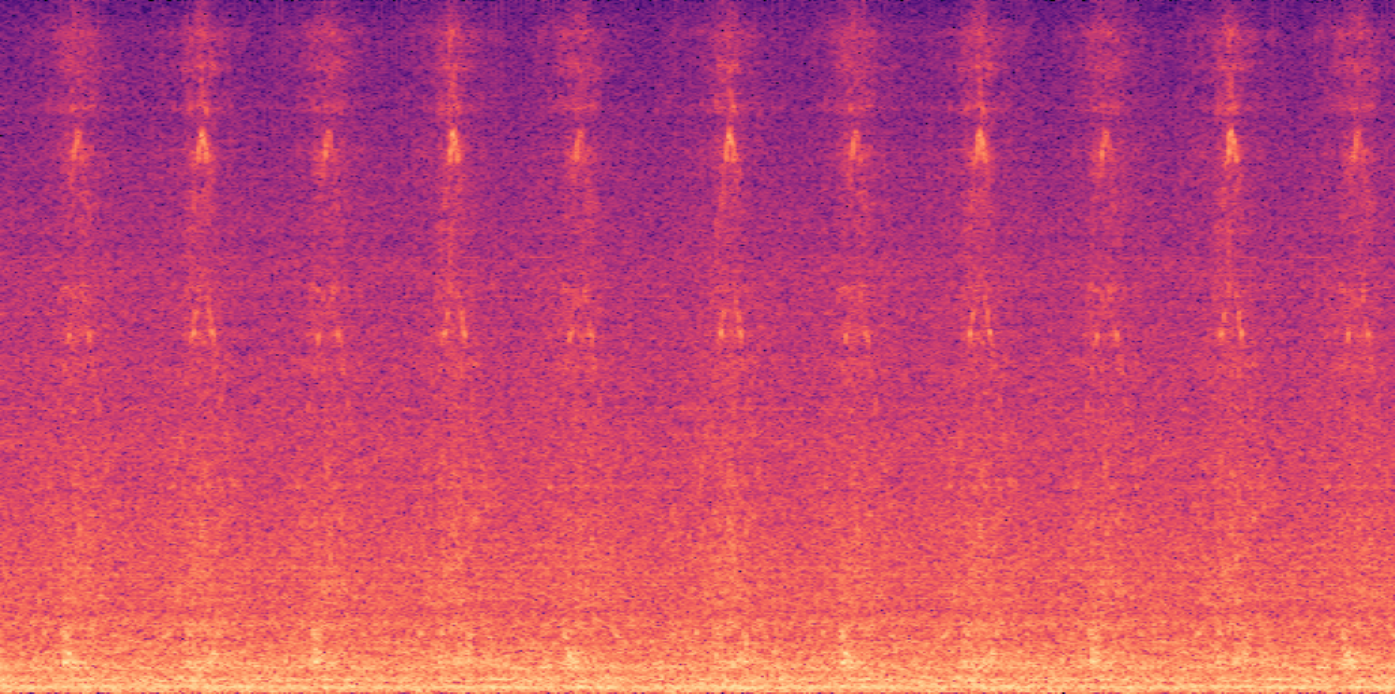}\hfill
\includegraphics[width=\mywidth]{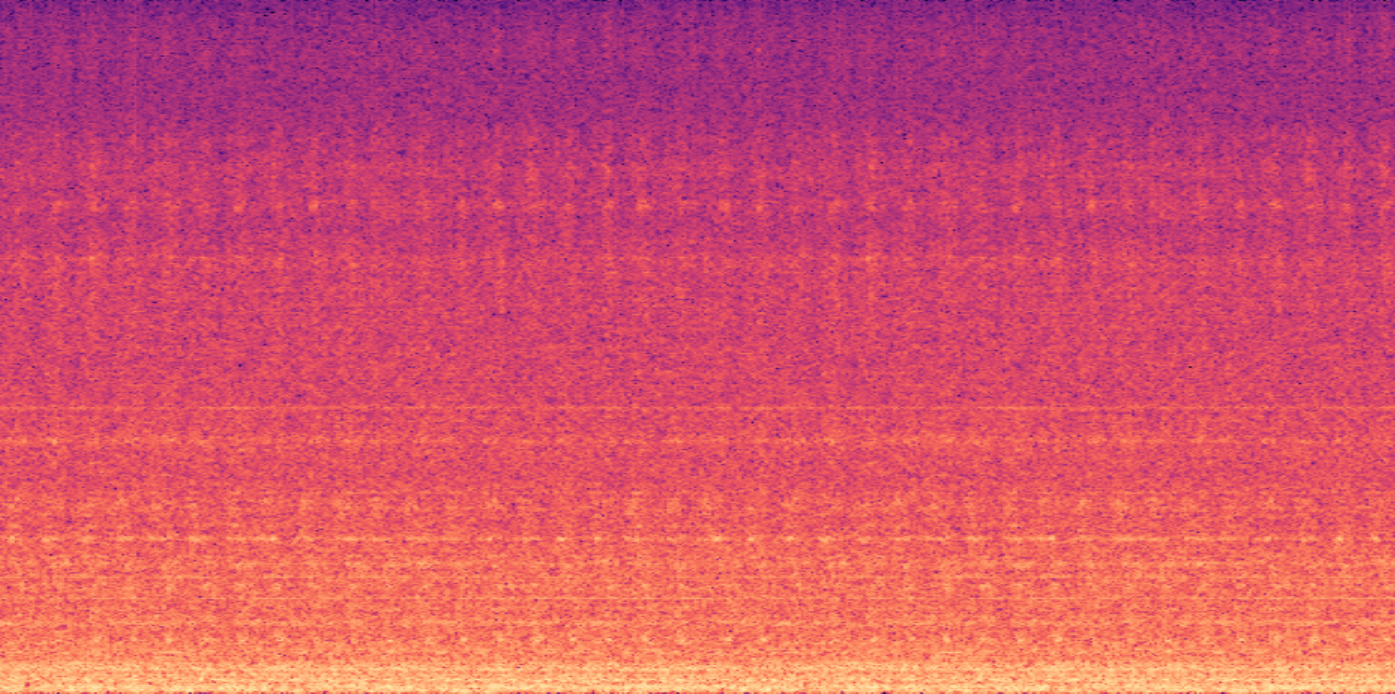}\hfill

\vspace{4pt}
\includegraphics[width=\mywidth]{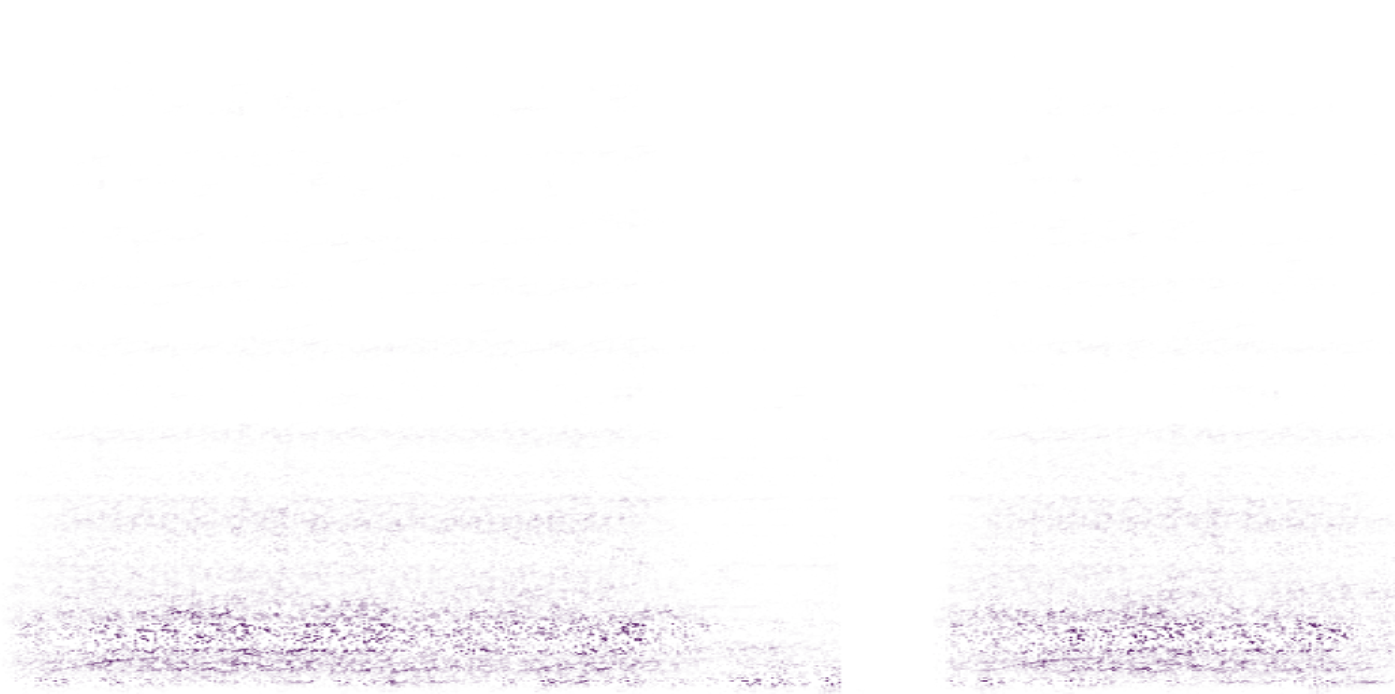}\hfill
\includegraphics[width=\mywidth]{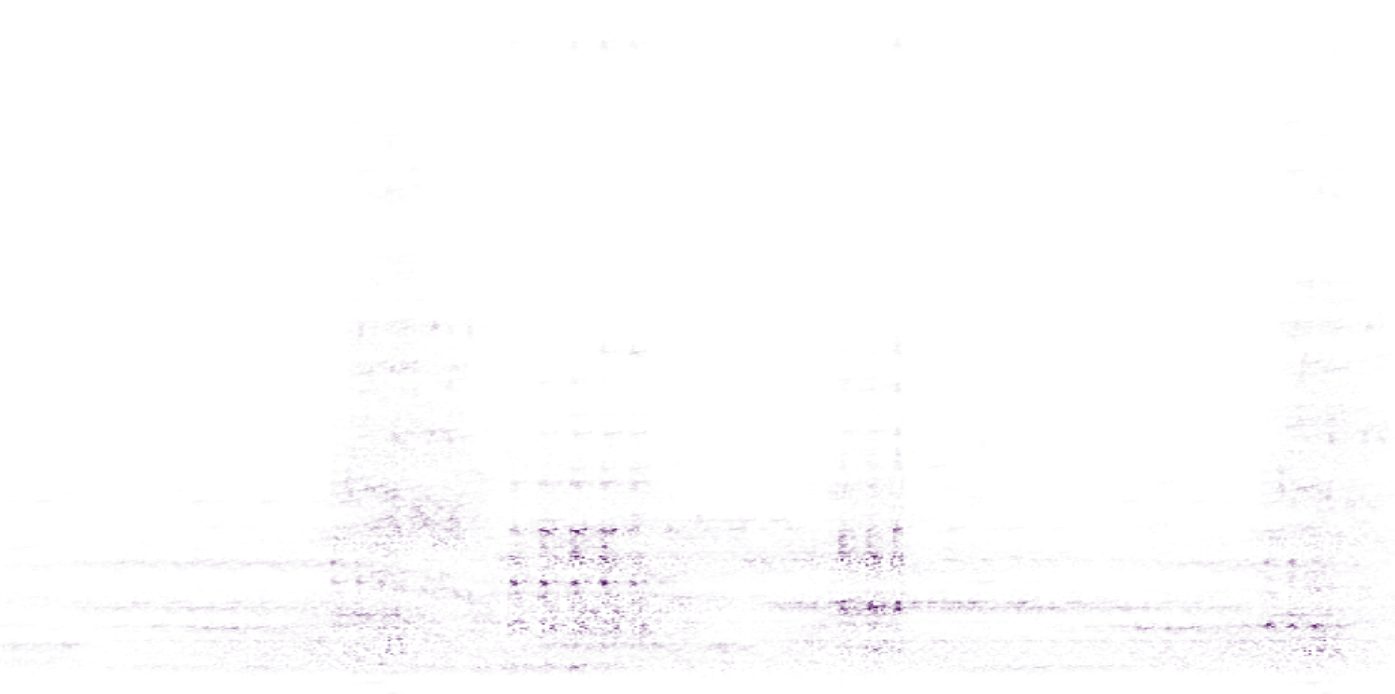}\hfill
\includegraphics[width=\mywidth]{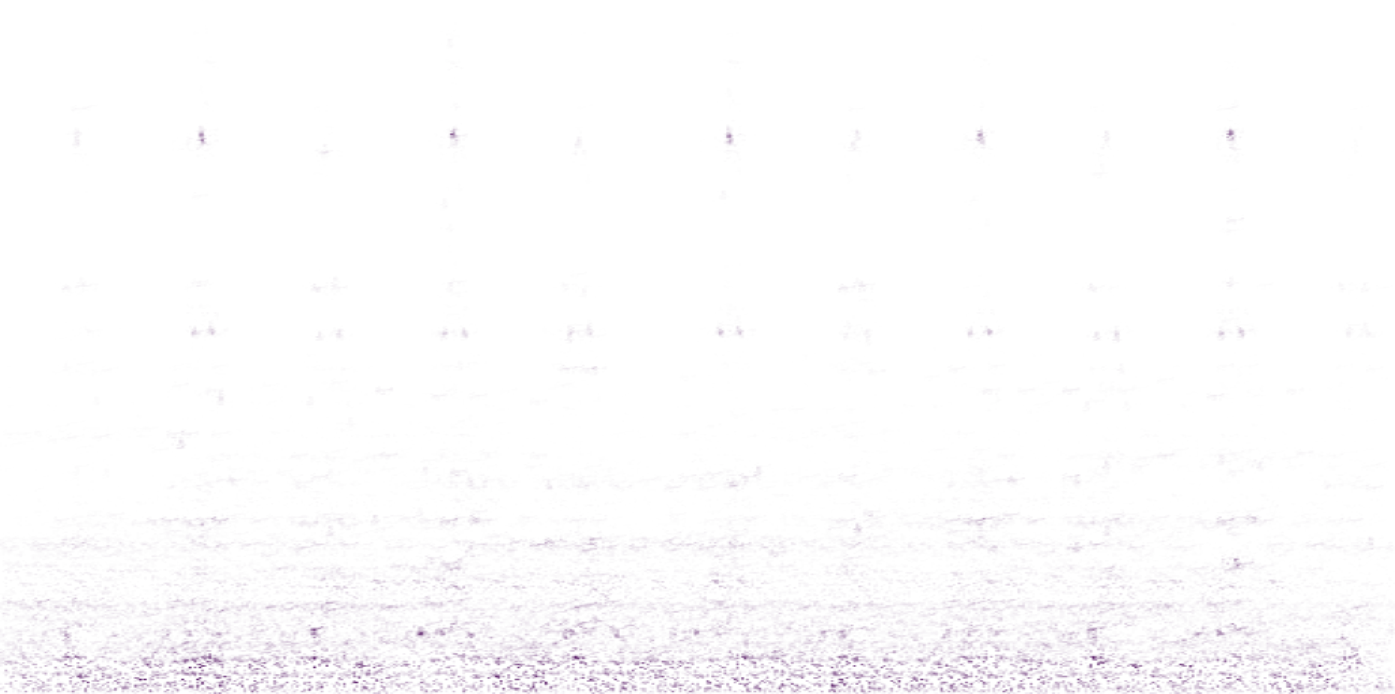}\hfill
\includegraphics[width=\mywidth]{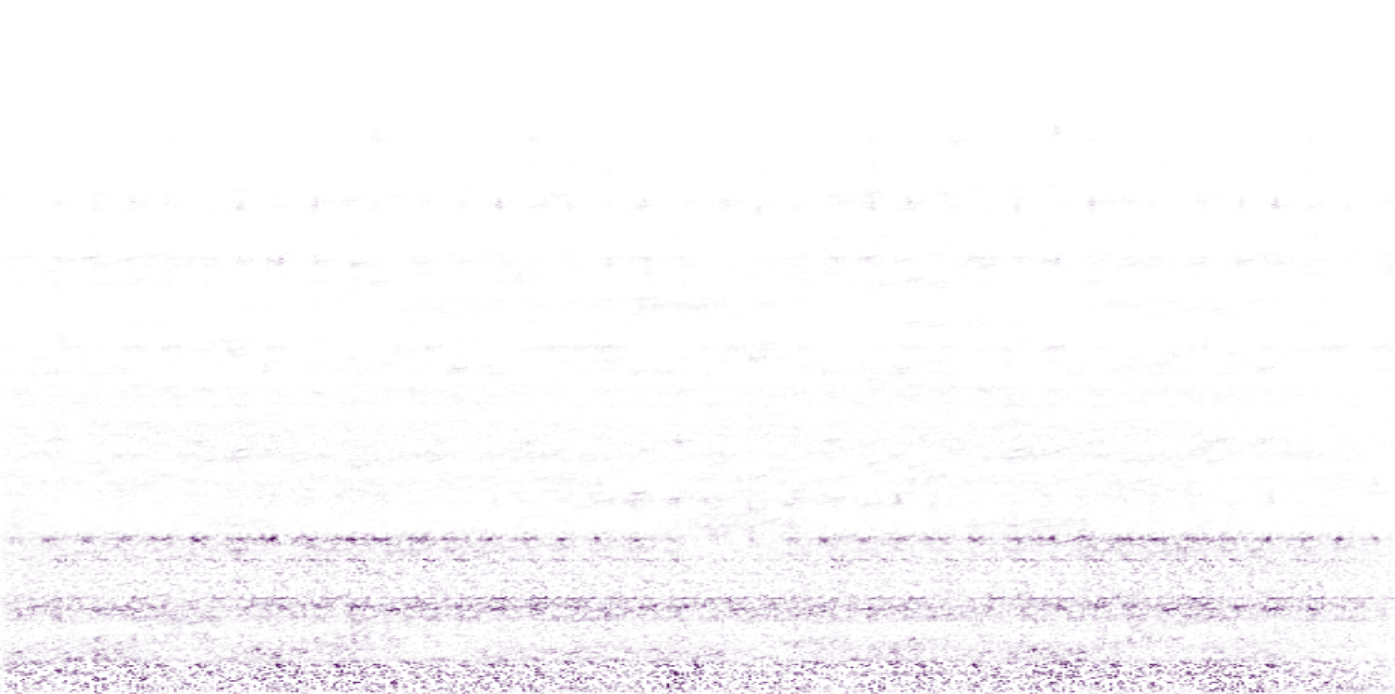}\hfill

\vspace{4pt}
\includegraphics[width=\mywidth]{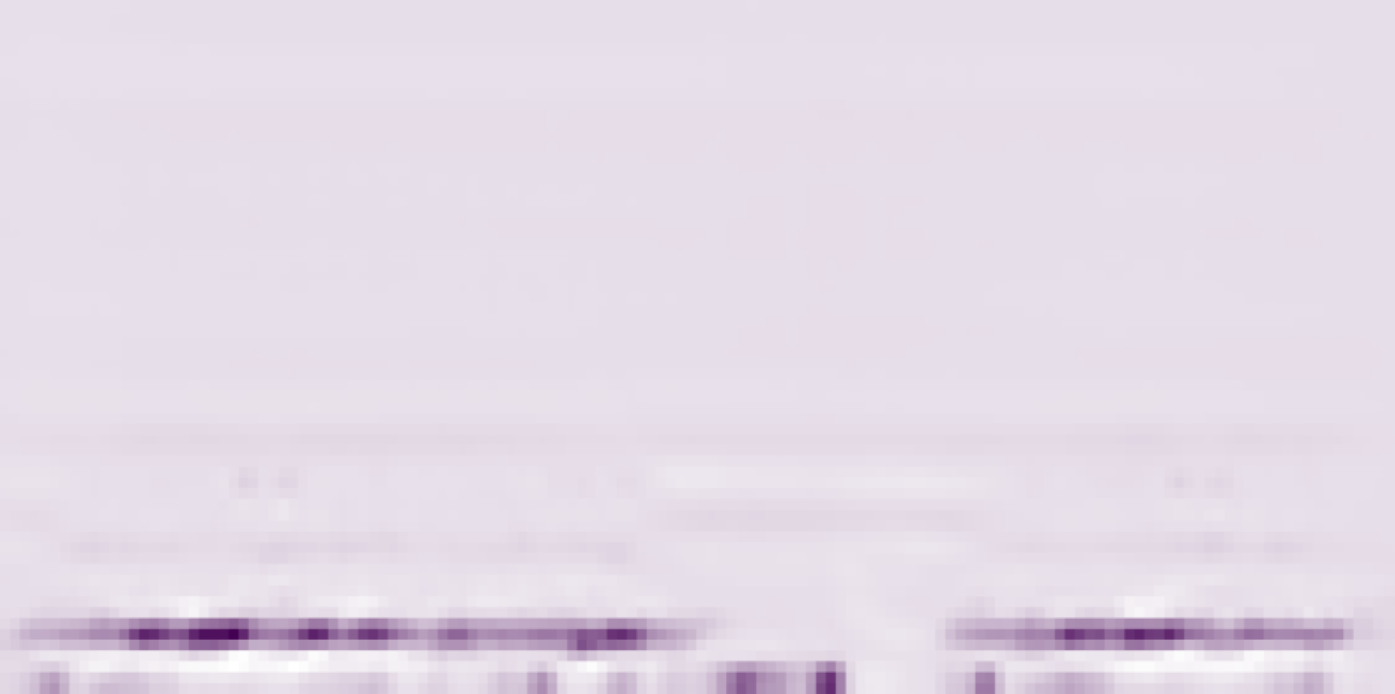}\hfill
\includegraphics[width=\mywidth]{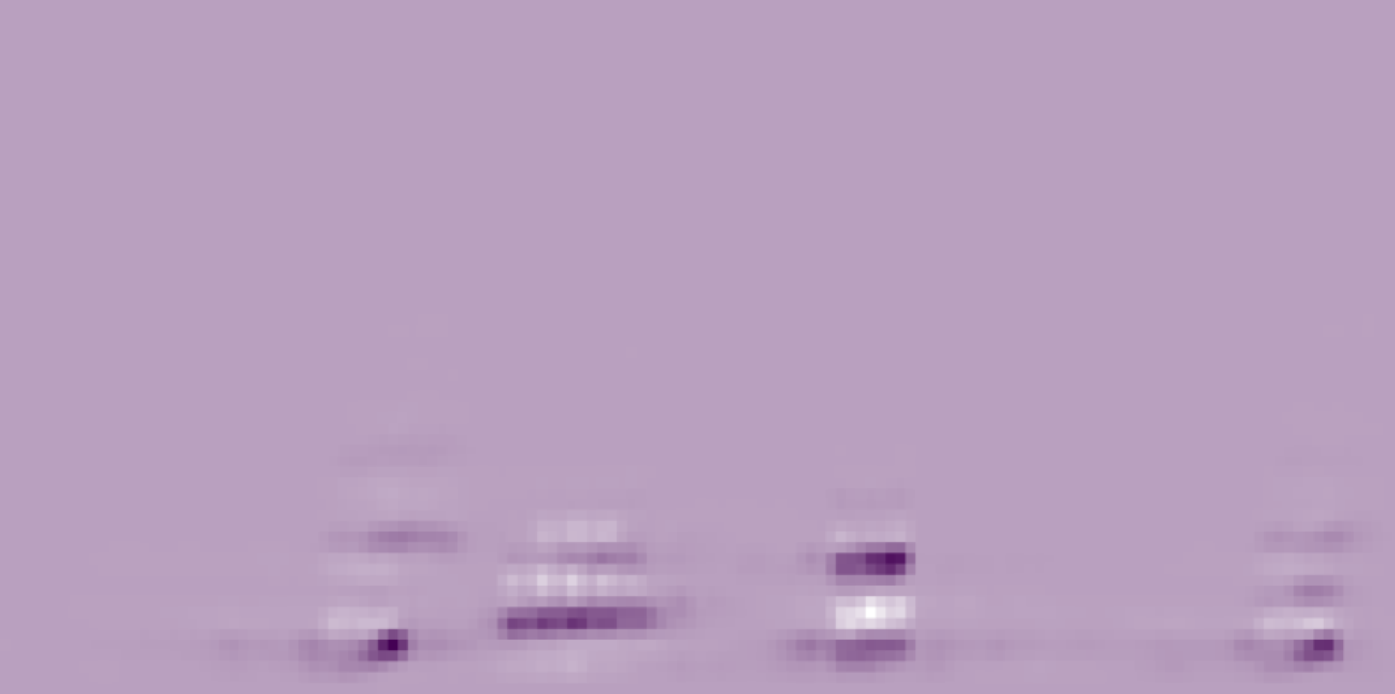}\hfill
\includegraphics[width=\mywidth]{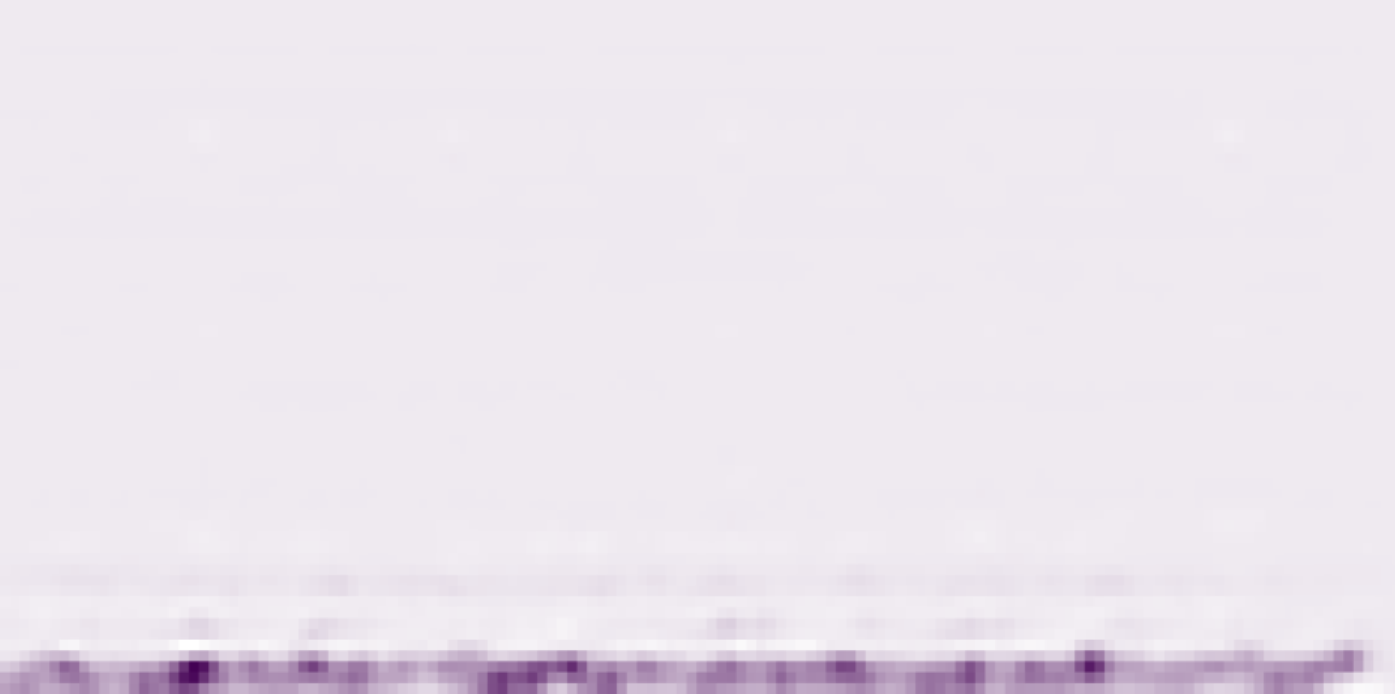}\hfill
\includegraphics[width=\mywidth]{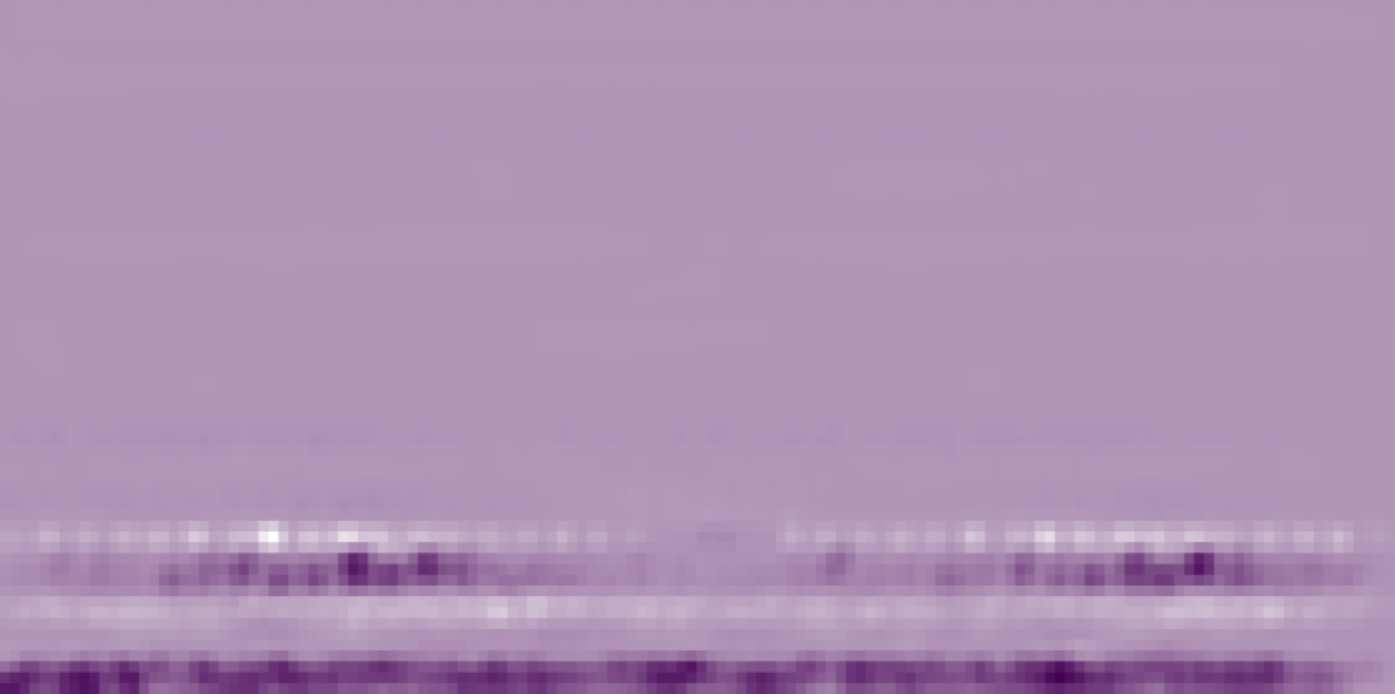}\hfill

\vspace{4pt}
\includegraphics[width=\mywidth]{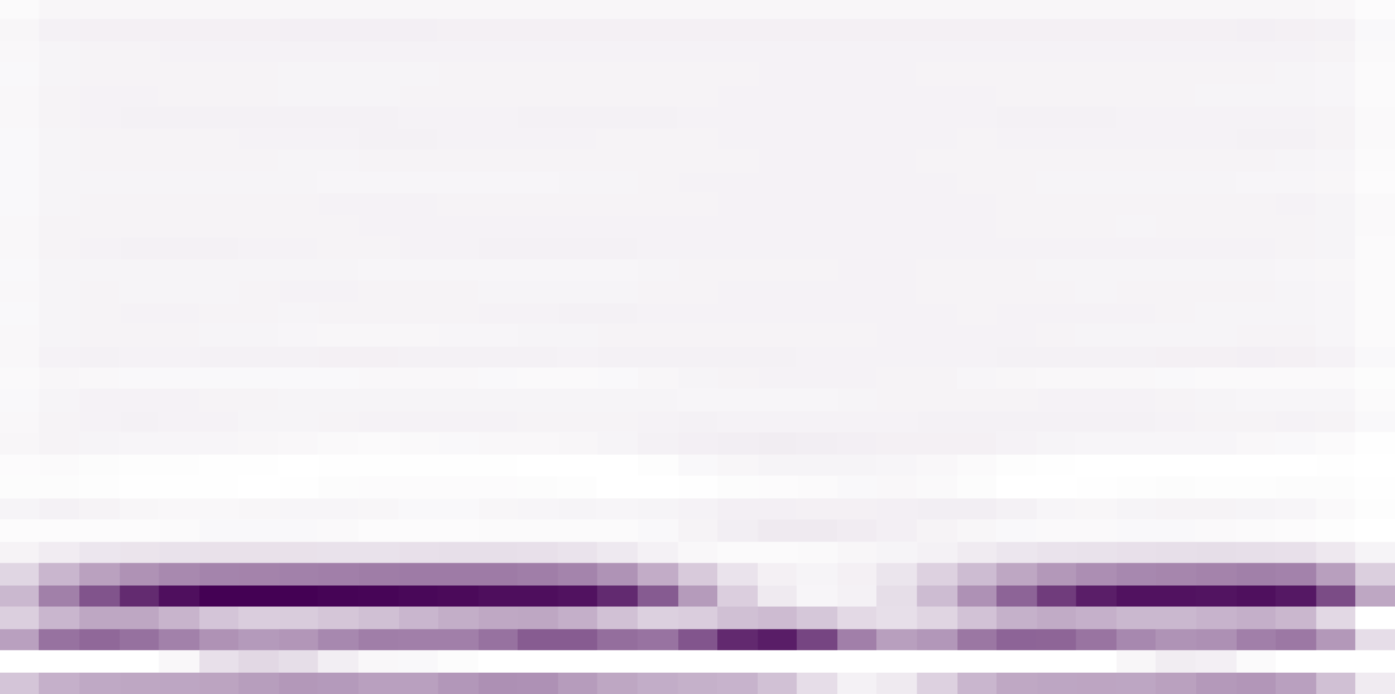}\hfill
\includegraphics[width=\mywidth]{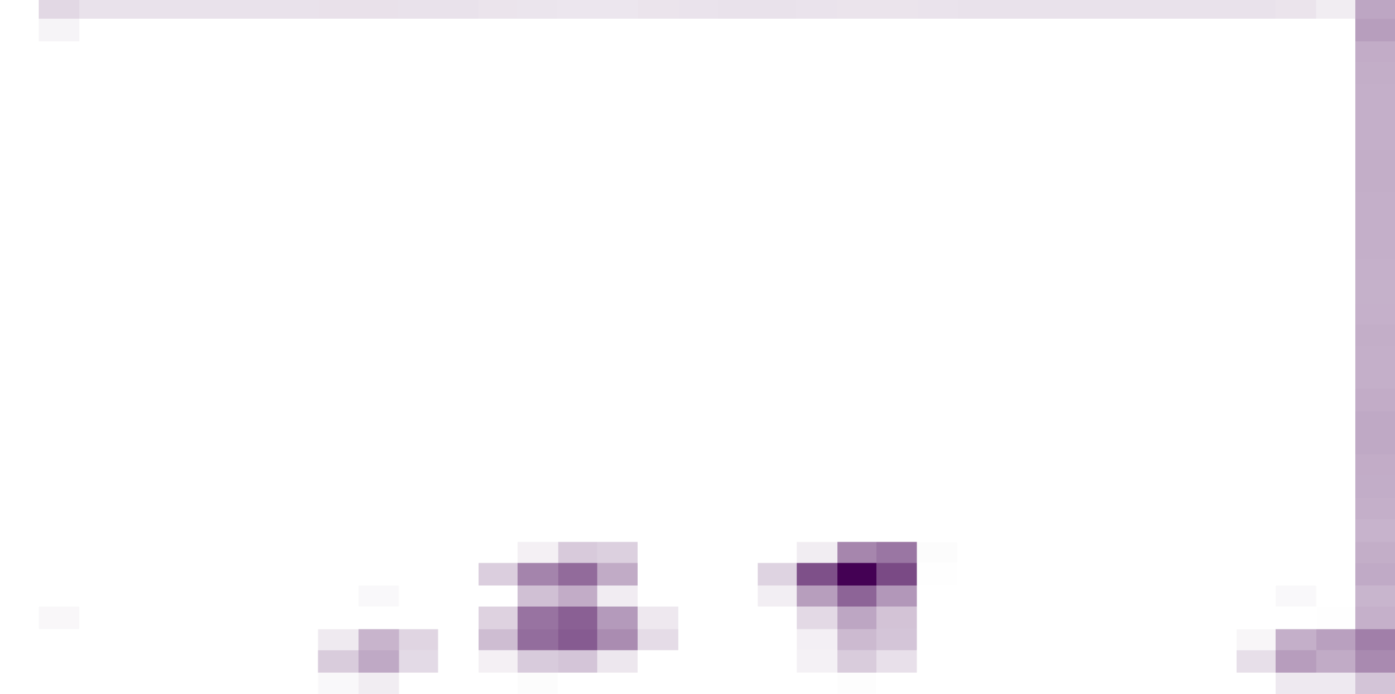}\hfill
\includegraphics[width=\mywidth]{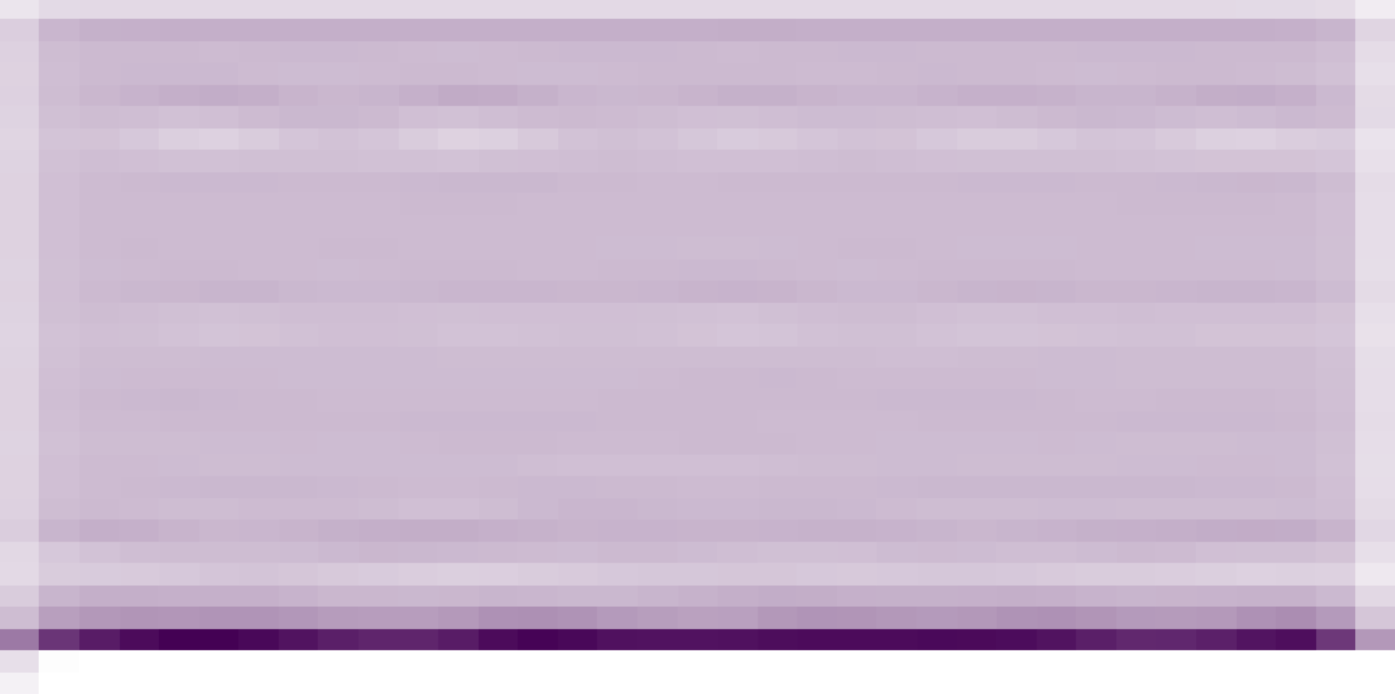}\hfill
\includegraphics[width=\mywidth]{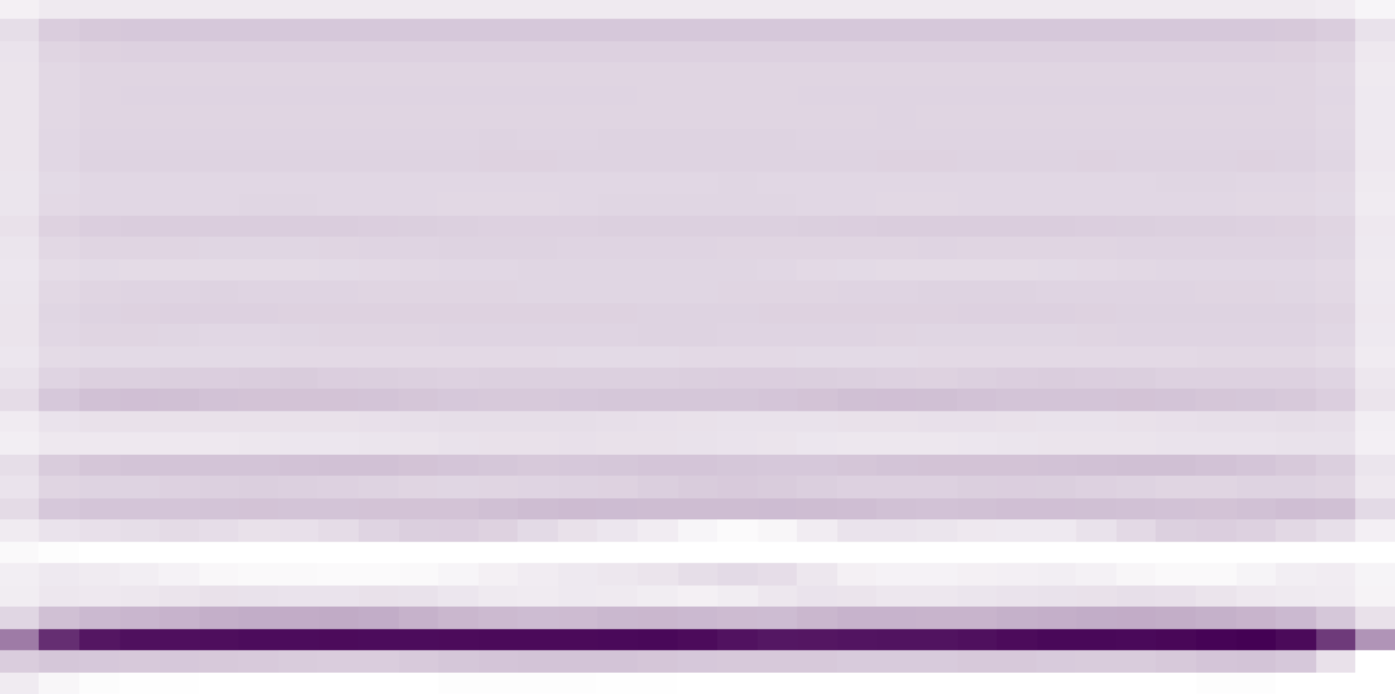}\hfill

\vspace{4pt}
\includegraphics[width=\mywidth]{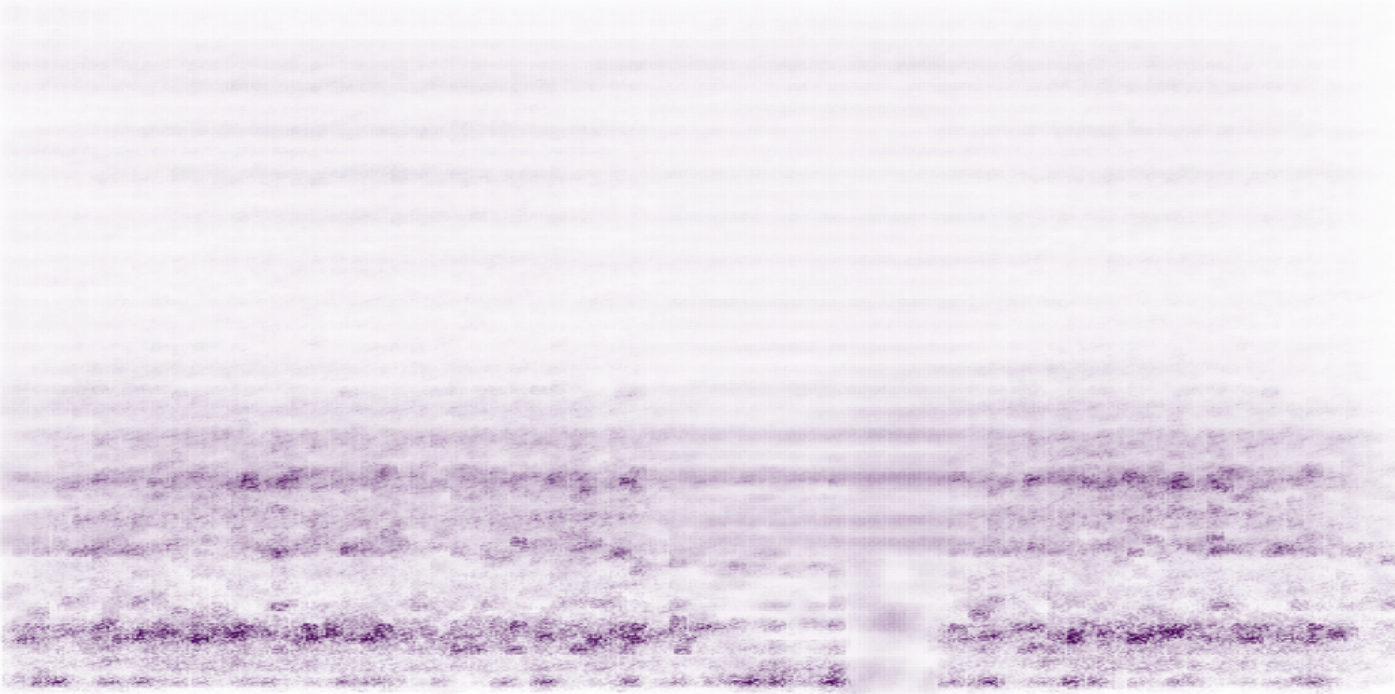}\hfill
\includegraphics[width=\mywidth]{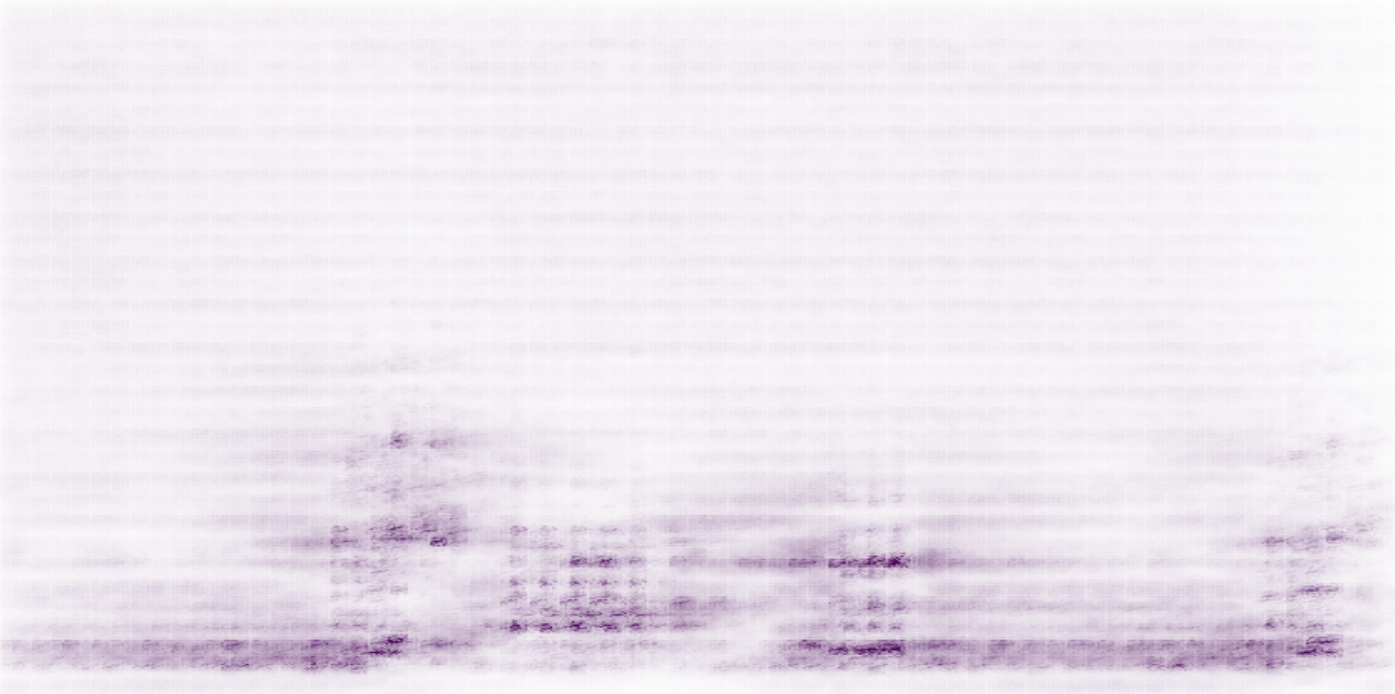}\hfill
\includegraphics[width=\mywidth]{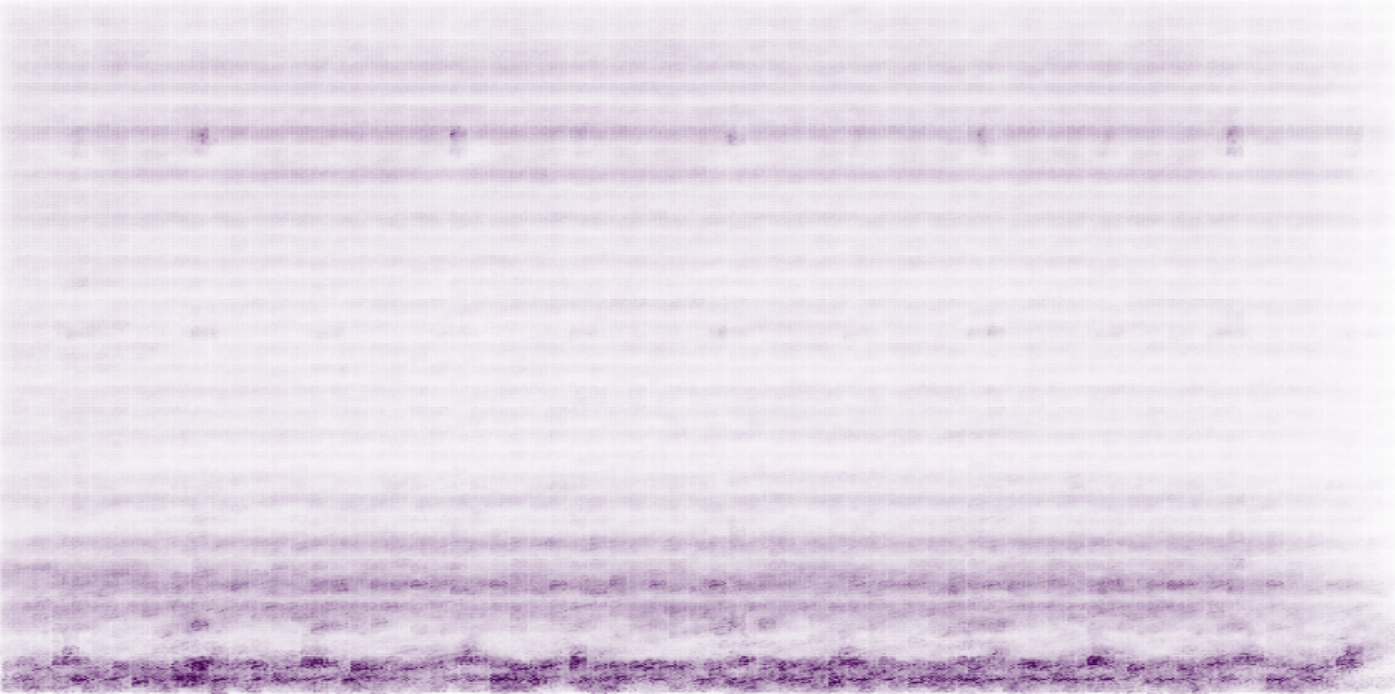}\hfill
\includegraphics[width=\mywidth]{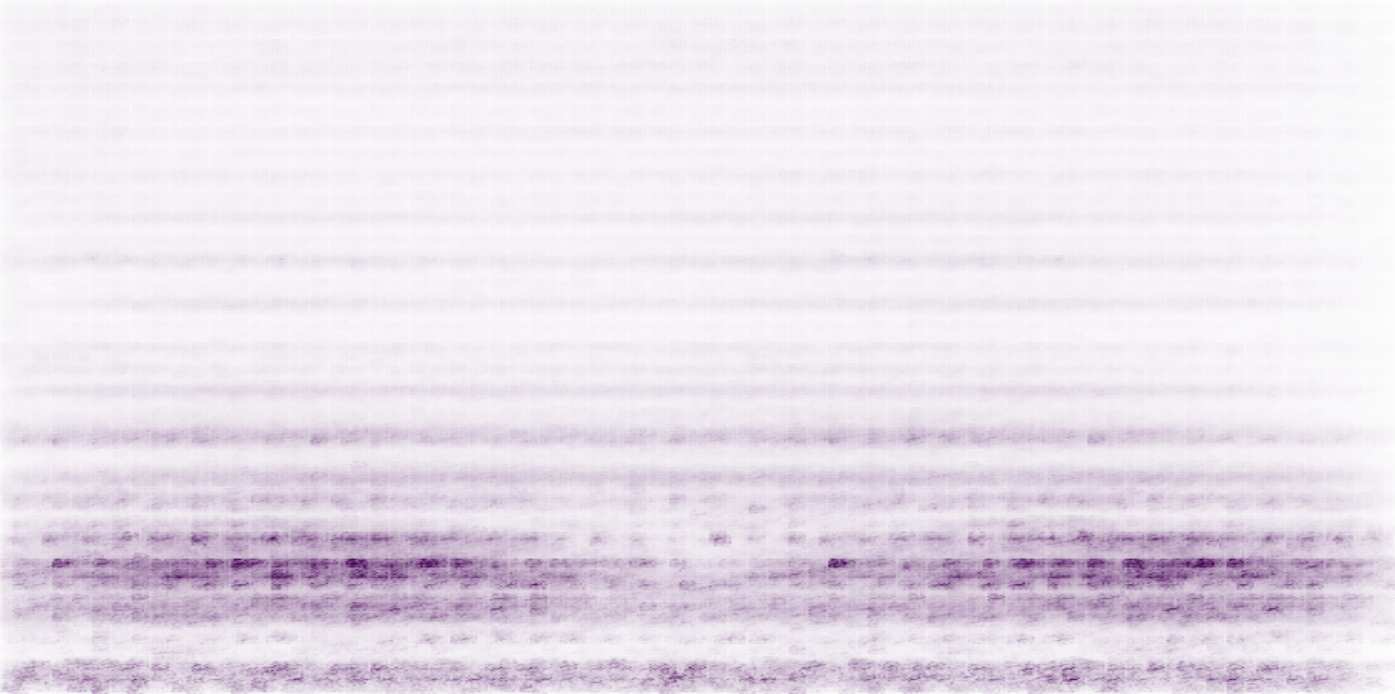}\hfill
\end{figure*}

Following the methodology of Section~\ref{sec:methodology 1}, the four attribution methods were applied to the input spectrograms to identify which time–frequency regions contributed most to the final classification. Figure~\ref{fig:five_images} shows the \gls{xai} attribution maps, with regions of higher relevance appearing in purple, while areas of low or negligible relevance are shown in white. Figure~\ref{fig:five_images} shows that across all methods, the attributions consistently concentrated on the low to mid frequency bands. These frequency ranges are typically the most energetic in the spectrograms, which may partly explain the strong focus observed.

\textbf{Integrated Gradients} produced the most fine-grained attribution patterns. These attributions closely followed the dominant spectral components and reflected sensitivity to small local variations, including rapid temporal fluctuations. Background regions were consistently assigned near-zero relevance, indicating that the method attributed virtually no importance to quieter areas of the input. This behaviour is expected, as Integrated Gradients accumulates the gradients along a path from a baseline to the input, causing regions with stronger activations and well-aligned gradients to dominate the attribution map.

\textbf{Occlusion} generated coarser relevance maps, highlighting broader regions rather than fine-grained details. Regions directly above or below important areas were often marked as unimportant, reflecting the block-wise perturbation used by the method. This is expected, as replacing larger patches with a constant value forces the model to respond to changes at a coarse spatial scale, producing broader relevance patterns and greater sensitivity to larger contextual regions.

\textbf{Grad-CAM} tended to assign low but non-zero relevance across most of the input and exhibited limited spatial precision. This lower spatial resolution prevented Grad-Cam from capturing features requiring short temporal context, particularly in samples containing fast-varying patterns. This is consistent with Grad-CAM’s reliance on the final convolutional layer’s feature maps, which are both spatially downsampled and semantically abstract, causing the resulting heatmaps to reflect general high-level activations rather than detailed spectral fluctuations.

\textbf{SmoothGrad} produced clearer and more differentiated relevance maps, with sharper contrasts between relevant and irrelevant areas. SmoothGrad highlighted a wider range of regions than the raw gradient-based method, suggesting improved stability and reduced noise. This is expected, as SmoothGrad averages attributions over many noisy perturbations of the input, effectively smoothing out gradient noise and reinforcing consistent, signal-dependent relevance patterns.

\section{Experiment 2: Frequency-Band Importance Analysis}
\label{sec:experiment 2}

To better understand which parts of the input the model depends on, this experiment systematically removes individual frequency bands from the spectrograms and measures the resulting change in anomaly detection performance, following the methodology of Section~\ref{sec:methodology 2}. Unlike \gls{xai} methods, which infer relevance from internal activations or gradients, this approach provides a direct behavioural test of the model’s reliance on specific spectral regions. By observing how performance shifts when certain bands are excluded, we can identify the frequencies that are truly critical for the model’s decisions. This creates a functional baseline against which \gls{xai} explanations can later be compared: if the model demonstrably needs a particular band to perform well, a reliable \gls{xai} method should highlight that same region as important.

\subsection{Frequency Band Removal Analysis}

\begin{table}[t]
    \centering
    \caption{Anomaly detection performance under frequency-band perturbation. Bold values indicate the best performance within each column.}
    \label{tab:freq_band_perturbation}
    \begin{tabular}{l *{4}{>{\raggedright\arraybackslash}p{1.1cm}}}
        \toprule
        Frequency Band	& Dev mean AUC &	Dev mean pAUC	 &Eval mean AUC	 &Eval mean pAUC\\
        \midrule
        Full Spectrum	 &73.5\%	 &57.2\%	 &\textbf{72\%}	 &\textbf{61.6\%}\\
        1	 &63.0\%	 &56.5\%	 &67.5\%	 &57.2\%\\
        2	 &68.0\%	 &57.3\%	 &68.9\%	 &60.1\%\\
        3	 &71.0\%	 &58.5\%	 &70.9\%	 &60.1\%\\
        4	 &70.9\%	 &59.2\%	 &66.9\%	 &58.2\%\\
        5	 &\textbf{76.8\%}	 &\textbf{59.3\%}	 &69.0\%	 &59.7\%\\
        \bottomrule

    \end{tabular}
\end{table}

Table \ref{tab:freq_band_perturbation} reports the anomaly detection performance when the model is trained and evaluated on either the full spectrum or modified spectrograms where individual frequency bands are removed. The frequency ranges correspond to successive \SI{1600}{\hertz} intervals: band 1 (\SIrange{0}{1600}{\hertz}), band 2 (\SIrange{1600}{3200}{\hertz}), band 3 (\SIrange{3200}{4800}{\hertz}), band 4 (\SIrange{4800}{6400}{\hertz}), and band 5 (\SIrange{6400}{8000}{\hertz}). Results are presented for both the development (Dev) and evaluation (Eval) sets using mean AUC and mean partial AUC (pAUC).

The full-spectrum model serves as the baseline, achieving mean AUC scores of 73.5\% (Dev) and 72\% (Eval), with corresponding pAUC values of 57.2\% and 61.6\%. Removing individual frequency bands leads to a range of effects, from substantial performance degradation to modest improvements, depending on the band and dataset split.

On the development set, removing the highest-frequency band (band 5) yields the strongest performance, with mean AUC increasing to 76.8\% and pAUC to 59.3\%. This suggests that, for these machines, very high frequencies may contain non-informative or noisy components that divert the model’s attention. However, this improvement does not transfer to the evaluation set, where the full spectrum performs best, indicating that the benefit of high-frequency removal is dataset- or machine-dependent rather than universally helpful.

In contrast, removing the lowest band (band 1) results in substantial performance loss, particularly on the development set where mean AUC falls to 63\%. This confirms that low-frequency information is essential for the model, consistent with the fact that many mechanical systems produce dominant energy in the low-frequency range through rotation, vibration, and motor activity. Excluding these components removes core information required to discriminate between normal and anomalous operation.

The intermediate bands (bands 2–4) produce mixed outcomes. Their removal occasionally yields small pAUC improvements, but overall performance remains below the full-spectrum baseline. This indicates that mid-frequency ranges provide useful information for anomaly detection.

This experiment shows which frequency regions the model truly relies on, providing a functional ground truth against which the \gls{xai} methods can later be evaluated. If a method accurately characterises model behaviour, it should assign relevance to the same bands whose removal harms performance, and downplay bands whose removal has little or even positive effect. This motivates the subsequent experiment, which examines whether the \gls{xai} explanations are consistent with the model’s demonstrated frequency dependencies.

\subsection{Machine-Level Frequency Sensitivity}

\begin{figure}[htbp]
    \centering
    \includegraphics[width=1\columnwidth]{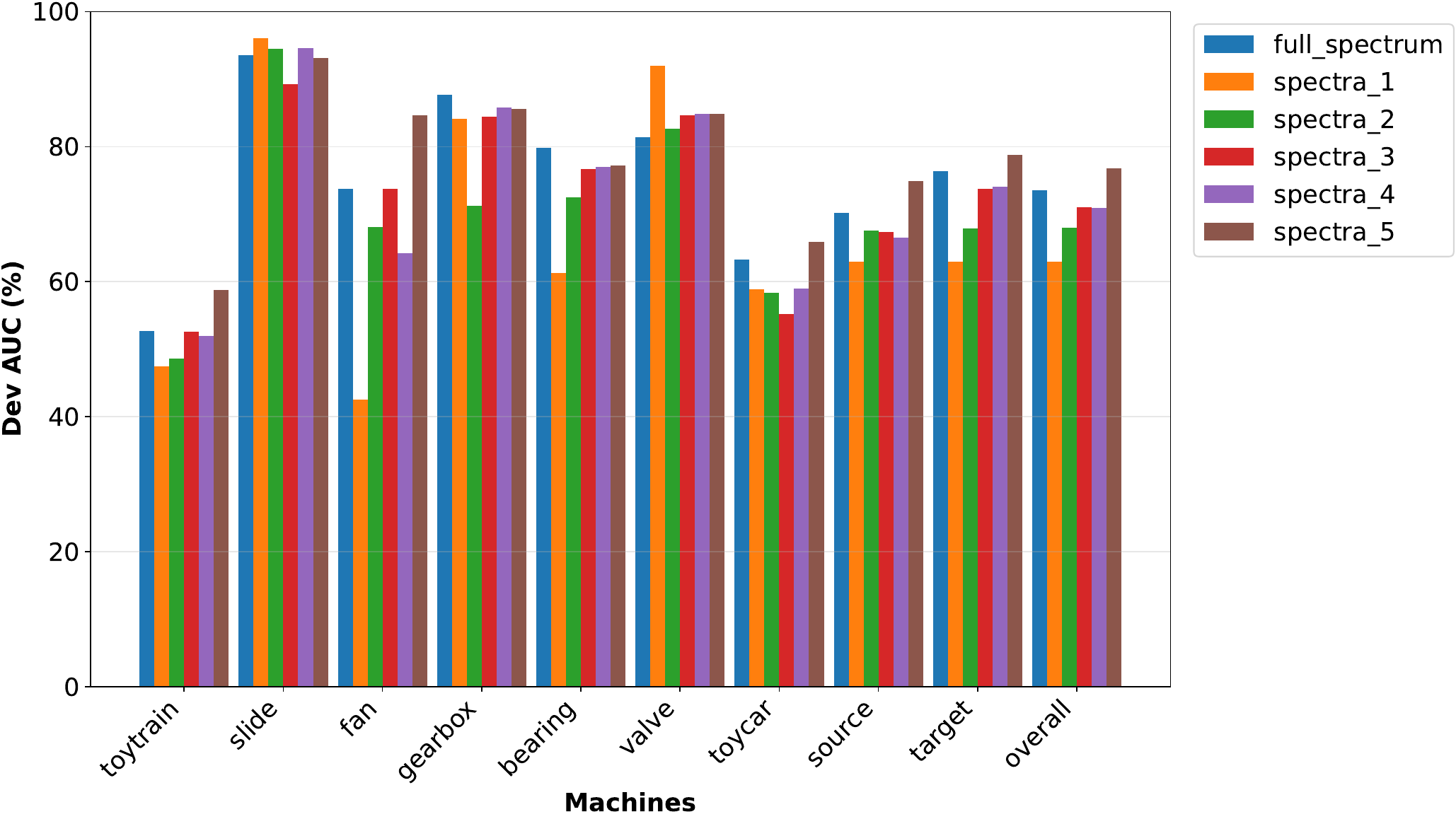}
    \caption{AUC results for each machine with different frequency bands on the development subset}
    \label{fig:Dev}
\end{figure}

\begin{figure}[htbp]
    \centering
    \includegraphics[width=1\columnwidth]{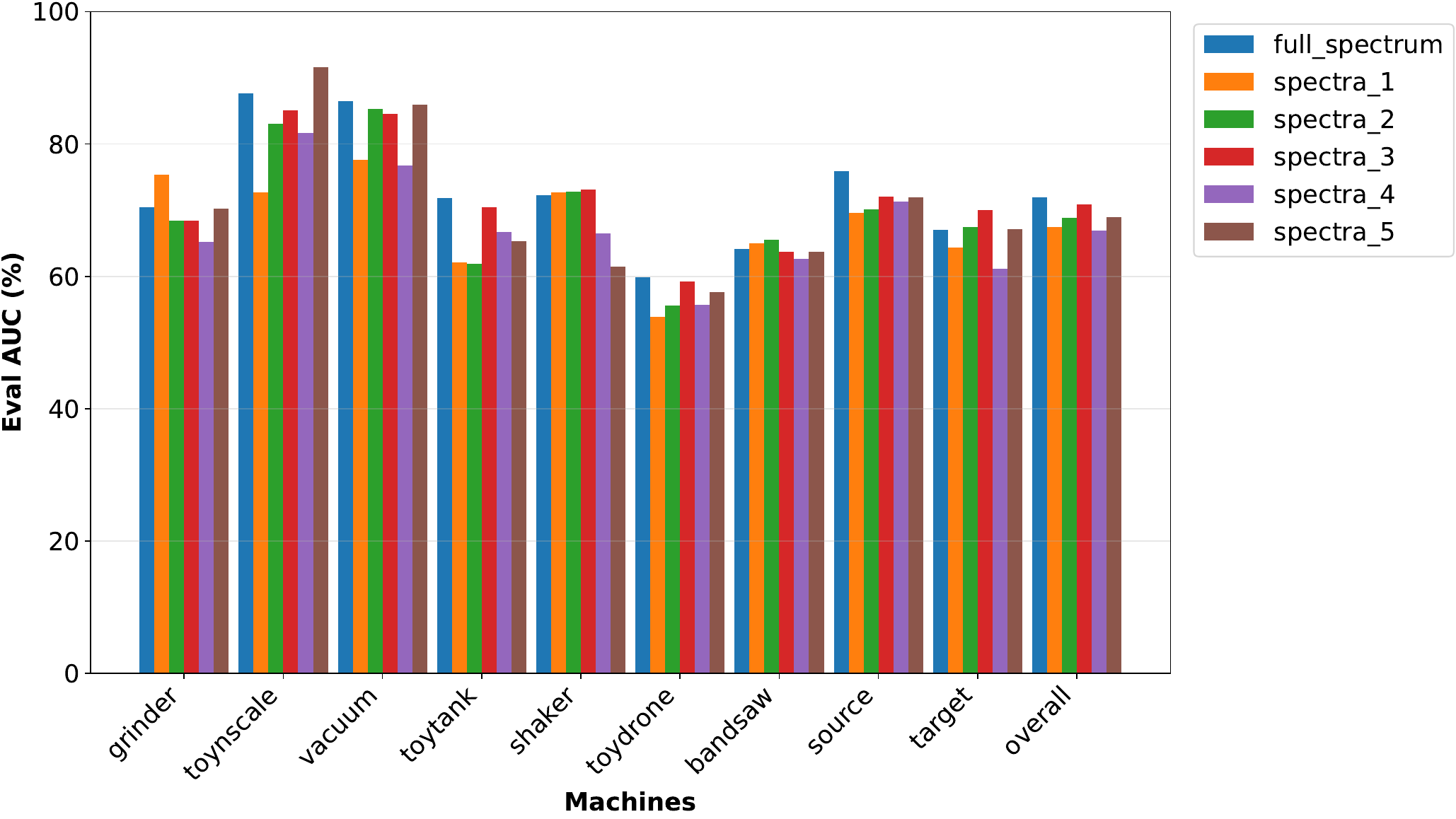}
    \caption{AUC results for each machine with different frequency bands on the evaluation subset}
    \label{fig:Eval}
\end{figure}

Figures \ref{fig:Dev} and \ref{fig:Eval} show the AUC scores for each machine type when removing individual frequency bands. Each coloured bar corresponds to one band removed (spectra 1 = \SIrange{0}{1600}{\hertz} removed, spectra 2 = \SIrange{1600}{3200}{\hertz} removed, etc.). The results highlight that frequency importance is strongly machine-dependent. For the \textit{valve}, removing the \SIrange{1600}{3200}{\hertz} band leads to substantial improvement, implying that although the model initially treated this band as informative, it contributed negatively to the model's predictions. For \textit{toytrain} and \textit{toynscale}, excluding the highest band (\SIrange{6400}{8000}{\hertz}) significantly improves performance, indicating that these machines contain high-frequency noise that distracts the model. Conversely, machines such as \textit{fan} and \textit{bearing} show sharp performance drops when the lowest band (\SIrange{0}{1600}{\hertz}) is removed, confirming that low-frequency acoustic signatures are essential for their anomaly patterns. Removing these frequencies eliminates fundamental information required for detecting abnormal behaviour. Other machines, such as \textit{slide} and \textit{bandsaw}, exhibit relatively stable performance across different band removals. This suggests that their discriminative features are spread more evenly across the spectrum, with no single band dominating the anomaly cues.

This experiment establishes a clear, machine-specific profile of how frequency removal affects prediction performance. This provides the empirical basis for Experiment 3, where the \gls{xai} methods are evaluated in terms of whether their relevance maps correctly reflect these model-specific frequency sensitivities.

\section{Experiment 3: Faithfulness Evaluation of XAI Methods}
\label{sec:experiment 3}

In this experiment, we follow the methodology of Section~\ref{sec:methodology 3} to quantify the relationship between (i) the mean relevance an \gls{xai} method assigns to a frequency band and (ii) the change in predicted anomaly score when that band is removed. This experiment evaluates the extent to which \gls{xai} methods in \gls{asd} reflect the spectral regions actually used by the model, comparing their attributions to changes in model predictions. Table~\ref{tab:xai_comparison} shows the Spearman correlation coefficients between mean relevance and $\Delta$Pred for each \gls{xai} method, with p-values shown in parentheses. 

\begin{table}
    \centering
    \caption{Spearman correlation between \gls{xai} relevance scores and model prediction changes under frequency-band removal. Cell colours encode correlation strength, with deeper red indicating weaker correlations and lighter yellow indicating stronger correlations. Higher values correspond to greater faithfulness, where relevance assigned by the \gls{xai} method aligns more closely with spectral regions that materially influence the model’s anomaly score. Values in brackets denote the corresponding $p$-values; correlations with $p$$<$0.05 are considered statistically significant.}
    \label{tab:xai_comparison}
    \begin{tabular}{l|*{4}{p{0.57in}}}
        \textbf{Machine} & \textbf{Integrated Gradients} & \textbf{Occlusion} & \textbf{Grad-CAM} & \textbf{SmoothGrad}\\
        \hline
bandsaw & \cellcolor{yellow!25!purple} 0.503 \textit{(0.006)} & \cellcolor{yellow!46!purple} 0.690 \textit{(0.000)} & \cellcolor{yellow!39!purple} 0.640 \textit{(0.120)} & \cellcolor{yellow!0!purple} 0.090 \textit{(0.000)}\\
bearing & \cellcolor{yellow!17!purple} 0.424 \textit{(0.000)} & \cellcolor{yellow!77!purple} 0.889 \textit{(0.000)} & \cellcolor{yellow!0!purple} -0.054 \textit{(0.188)} & \cellcolor{yellow!9!purple} 0.311 \textit{(0.000)}\\
fan & \cellcolor{yellow!14!purple} 0.389 \textit{(0.000)} & \cellcolor{yellow!94!purple} 0.976 \textit{(0.000)} & \cellcolor{yellow!5!purple} 0.239 \textit{(0.162)} & \cellcolor{yellow!8!purple} 0.292 \textit{(0.000)}\\
gearbox & \cellcolor{yellow!26!purple} 0.518 \textit{(0.003)} & \cellcolor{yellow!34!purple} 0.595 \textit{(0.002)} & \cellcolor{yellow!16!purple} 0.418 \textit{(0.000)} & \cellcolor{yellow!10!purple} 0.332 \textit{(0.000)}\\
grinder & \cellcolor{yellow!88!purple} 0.949 \textit{(0.000)} & \cellcolor{yellow!67!purple} 0.827 \textit{(0.000)} & \cellcolor{yellow!32!purple} 0.577 \textit{(0.165)} & \cellcolor{yellow!14!purple} 0.383 \textit{(0.000)}\\
shaker & \cellcolor{yellow!19!purple} 0.449 \textit{(0.770)} & \cellcolor{yellow!94!purple} 0.974 \textit{(0.000)} & \cellcolor{yellow!32!purple} 0.579 \textit{(0.000)} & \cellcolor{yellow!13!purple} 0.371 \textit{(0.077)}\\
slider & \cellcolor{yellow!22!purple} 0.474 \textit{(0.000)} & \cellcolor{yellow!88!purple} 0.943 \textit{(0.000)} & \cellcolor{yellow!60!purple} 0.781 \textit{(0.186)} & \cellcolor{yellow!25!purple} 0.502 \textit{(0.034)}\\
ToyCar & \cellcolor{yellow!22!purple} 0.480 \textit{(0.000)} & \cellcolor{yellow!75!purple} 0.878 \textit{(0.000)} & \cellcolor{yellow!84!purple} 0.922 \textit{(0.000)} & \cellcolor{yellow!4!purple} 0.225 \textit{(0.000)}\\
ToyDrone & \cellcolor{yellow!14!purple} 0.381 \textit{(0.000)} & \cellcolor{yellow!27!purple} 0.529 \textit{(0.127)} & \cellcolor{yellow!60!purple} 0.788 \textit{(0.000)} & \cellcolor{yellow!11!purple} 0.341 \textit{(0.000)}\\
ToyNscale & \cellcolor{yellow!34!purple} 0.595 \textit{(0.861)} & \cellcolor{yellow!39!purple} 0.639 \textit{(0.024)} & \cellcolor{yellow!0!purple} -0.154 \textit{(0.000)} & \cellcolor{yellow!46!purple} 0.680 \textit{(0.057)}\\
ToyTank & \cellcolor{yellow!34!purple} 0.592 \textit{(0.000)} & \cellcolor{yellow!84!purple} 0.921 \textit{(0.000)} & \cellcolor{yellow!79!purple} 0.898 \textit{(0.003)} & \cellcolor{yellow!5!purple} 0.249 \textit{(0.000)}\\
ToyTrain & \cellcolor{yellow!13!purple} 0.380 \textit{(0.000)} & \cellcolor{yellow!92!purple} 0.961 \textit{(0.000)} & \cellcolor{yellow!0!purple} -0.253 \textit{(0.008)} & \cellcolor{yellow!25!purple} 0.509 \textit{(0.000)}\\
Vacuum & \cellcolor{yellow!12!purple} 0.367 \textit{(0.000)} & \cellcolor{yellow!92!purple} 0.968 \textit{(0.000)} & \cellcolor{yellow!36!purple} 0.605 \textit{(0.000)} & \cellcolor{yellow!10!purple} 0.320 \textit{(0.000)}\\
valve & \cellcolor{yellow!12!purple} 0.353 \textit{(0.000)} & \cellcolor{yellow!51!purple} 0.726 \textit{(0.000)} & \cellcolor{yellow!4!purple} 0.209 \textit{(0.002)} & \cellcolor{yellow!56!purple} 0.751 \textit{(0.542)}\\
overall & \cellcolor{yellow!27!purple} 0.530 \textit{(0.000)} & \cellcolor{yellow!77!purple} 0.884 \textit{(0.000)} & \cellcolor{yellow!28!purple} 0.536 \textit{(0.000)} & \cellcolor{yellow!16!purple} 0.400 \textit{(0.000)}\\
    \end{tabular}
\end{table}

\textbf{Integrated Gradients} shows moderate but reliable faithfulness across machines, with an overall correlation of 0.530. Stronger performance was observed on machines with stable narrow-band spectral structures (e.ge, grinder, gearbox).

\textbf{Occlusion} is consistently the most faithful method, achieving the highest correlation for nearly all machines and a dominant overall average correlation of 0.884. This strongly suggests that direct perturbation-based relevance estimates align most closely with the model’s true decision process. However, it is worth noting that Occlusion’s exceptionally strong performance is partly expected given how closely the functional ground truth aligns with Occlusion's underlying mechanism. Occlusion estimates feature importance by systematically deleting small time–frequency patches and measuring the resulting change in the model’s output. The frequency-band removal experiment follows the same principle. By perturbing spectral regions and observing prediction shifts, the evaluation procedure inherently mirrors the way Occlusion generates relevance. As a result, Occlusion is naturally advantaged in this setting, and its high correlations reflect not only its faithfulness but also the structural similarity between Occlusion and the functional ground truth.

\textbf{Grad-CAM} exhibits highly variable behaviour, ranging from strong correlations (e.g., ToyCar, ToyTank) to negative or insignificant correlations (e.g., bearing, ToyNscale, ToyTrain).
This indicates that Grad-CAM may not consistently capture frequency-localised mechanisms in audio spectrogram models.

\textbf{SmoothGrad} generally underperforms, achieving the lowest overall correlation (0.400) and showing instability across machines. This suggests that noise-based smoothing may obscure spectral precision, making this smoothing less suitable for \gls{asd} relevance estimation.

\section{Discussion}

This work examines the faithfulness with which common \gls{xai} methods explain spectrogram-based machine sound anomaly detection models. Through qualitative attribution analysis, frequency-band perturbation experiments, and correlation-based evaluation of attribution faithfulness, we observe consistent patterns in both model behaviour and the extent to which different \gls{xai} methods capture that behaviour. 

\subsection{Qualitative Analysis of Attribution Methods}

The qualitative comparisons in Experiment 1 showed that all four attribution methods broadly emphasise low- and mid-frequency components, but differ in granularity, stability, and selectivity. Integrated Gradients produced highly detailed saliency patterns, Occlusion highlighted coarse contextual regions, SmoothGrad sharpened gradient-based explanations at the expense of some spectral precision, and Grad-CAM yielded diffuse maps with limited frequency localisation. This diversity reflects both the underlying mechanisms of each \gls{xai} method and the well-known problem that attribution methods routinely disagree. The findings reinforce that visually plausible heatmaps should not be assumed to reflect the model’s true dependencies.

\subsection{Frequency-Removal Analysis and Model Sensitivity}

Experiment 2 established a functional ground truth by quantifying how the model’s anomaly score changes when entire frequency bands are removed. Across machine types, low-frequency energy (\SIrange{0}{1600}{\hertz}) consistently contributed the most to model predictions, as its removal produced the largest performance decreases. This aligns with the physical characteristics of many mechanical systems, where rotation, vibration, and harmonic motor behaviour primarily manifest in lower frequency bands.

Higher frequencies were less informative, and in some cases, removal of the \SIrange{6400}{8000}{\hertz} band led to improved performance for machines such as ToyTrain, ToyNscale, and ToyCar, suggesting that high-frequency noise may introduce irrelevant cues. Mid-frequency bands exhibited machine-dependent relevance, highlighting that \gls{asd} models learn spectral biases specific to the acoustic properties of each device.

These machine-specific frequency dependencies provide an objective baseline for evaluating \gls{xai} explanations: reliable methods should emphasise the same bands whose removal the model is most sensitive to.

\subsection{Faithfulness Evaluation of XAI Methods}

The results of Experiment 3 show that different \gls{xai} methods vary in their ability to reflect the model’s reliance on specific spectral regions. Occlusion consistently showed the highest correlation between attributions and the change in anomaly score under frequency-band removal, at both the machine-specific and overall level. This correspondence is expected because Occlusion and the functional ground truth both rely on local input perturbations and measure resulting output changes, providing a close alignment between the explanation method and the functional ground truth.

Integrated Gradients showed moderate correlation values across machines, accurately capturing essential frequency dependencies in many cases. Its detailed attribution patterns and completeness properties contributed to recovering narrow-band spectral structures. However, reliance on gradient information introduces sensitivity to baseline selection and input noise. This reliance limits correlation performance for some machine types.

Grad-CAM exhibited variable correlation across machines. Some machines showed reasonable alignment with model sensitivity, while others produced low or negative correlation values. This indicates that Grad-CAM explanations can differ from the actual frequency importance. This behaviour is consistent with Grad-CAM’s reliance on the spatial structure of convolutional feature maps, which may not align with frequency-localised cues. SmoothGrad produced visually stable saliency maps but lower correlation with the functional ground truth. The smoothing process reduces the fidelity of fine-grained spectral information. As a result, SmoothGrad explanations show weaker correspondence with perturbation-based sensitivity.

\subsection{Implications for interpreting and designing ASD models}

The combined evidence from the three experiments has several important implications for the use and interpretation of \gls{xai} in machine \gls{asd}. First, the difference between qualitative heatmaps and quantitative faithfulness highlights the need for caution when interpreting visual explanations in spectrogram-based systems. An attribution method that appears intuitive or aesthetically clear may not reflect the cues the model is truly exploiting. This is particularly relevant in industrial applications, where misleading explanations could lead to incorrect assessments of machine condition.

The frequency-removal results also show that \gls{asd} models exhibit strong biases toward specific spectral regions, especially low-frequency bands, which often contain the fundamental mechanical signatures of normal and abnormal operation. High-frequency content, by contrast, can introduce noise that distracts the model rather than contributing meaningful information. Incorporating band-level diagnostics into the model development process therefore provides a reliable way to identify harmful spectral biases and to guide feature-engineering choices such as filtering strategies or frequency-weighted losses. Models that appear to perform well may in fact rely on narrow spectral regions that are unstable across environments or recording devices, and the proposed framework offers a means to uncover these tendencies.

From a practical perspective, the findings suggest that perturbation-based methods, despite their computational expense, are currently the most reliable tools for interpreting spectrogram-based \gls{asd} models. In settings where faithful explanations are critical, such as fault diagnosis, maintenance planning, or safety auditing, Occlusion-style analyses may be preferable to gradient-based alternatives. Conversely, methods such as Grad-CAM and SmoothGrad should be used with caution in audio, as they may not reveal the mechanisms that truly drive model predictions. Ultimately, the results underscore the importance of combining domain knowledge about machine physics with rigorous model-centric analysis when deploying explainable \gls{asd} systems.

\subsection{Limitations and opportunities for future research}

Although this study provides the first systematic evaluation of \gls{xai} faithfulness for machine \gls{asd}, several important limitations point toward promising directions for future research.

\textbf{Limited Temporal Scope of Ablation Analysis} The current analysis focuses exclusively on frequency-based perturbations, leaving temporal masking and joint time–frequency interaction effects unexplored. Real mechanical faults often manifest through transient events or evolving patterns across time, and extending the perturbation framework to include temporal structure could provide a more comprehensive understanding of model sensitivity.

\textbf{Lack of Perceptually Motivated Frequency Representations} Another limitation lies in the use of linear frequency bands rather than perceptually informed scales. While the goal of this work is to evaluate model-centric faithfulness rather than human interpretability, integrating psychoacoustic principles, such as Bark or ERB scaling \cite{zwlcker1961subdivision, PMID:6630731}, may yield explanations that better align with how domain experts reason about sound. This alignment could be especially valuable in industrial settings where human operators must interpret model behaviour in conjunction with their own diagnostic expertise.

\textbf{Architectural Specificity of the Evaluation Framework} The study also focuses on a single convolutional encoder architecture inspired by the FeatEx approach. As \gls{asd} models increasingly shift toward transformer-based or diffusion-based architectures with richer attention mechanisms, it will be necessary to revisit how \gls{xai} methods behave under these new paradigms and whether perturbation-based faithfulness evaluations require adaptation. Additionally, robustness under domain shift remains an open question. The faithfulness of an explanation method may depend not only on the architecture but also on the data distribution, background noise, or environment in which the model is deployed.

\textbf{Absence of Human-Centred and Operational Validation} While this work evaluates faithfulness at the model level, future research may explore how explanations can be integrated into human-in-the-loop workflows. Aligning attribution maps with known fault signatures such as harmonics, sidebands, or modulation patterns could improve the practical usefulness of \gls{xai} for maintenance engineers and acoustic diagnosticians.

\section{Conclusion}

This work introduced a new quantitative framework for evaluating the faithfulness of \gls{xai} methods in machine \gls{asd}, addressing a gap in existing audio explainability research, which has relied almost entirely on qualitative inspection of attribution maps. The proposed approach links \gls{xai} relevance directly to the model’s behavioural sensitivity by correlating attribution scores with prediction changes under systematic frequency-band removal. This provides, for the first time, an objective criterion for determining whether an explanation method truly reflects the spectral information that drives an \gls{asd} model’s decisions.

Across extensive experiments on the DCASE 2023 Task 2 dataset, the results demonstrate that \gls{xai} methods differ in their faithfulness. Occlusion exhibits the strongest and most consistent alignment with model behaviour, Integrated Gradients shows moderate reliability, and gradient-visualisation methods such as Grad-CAM and SmoothGrad often fail to capture spectral dependencies. The frequency-removal analysis further reveals that \gls{asd} models rely heavily on low-frequency energy while high-frequency content can introduce misleading cues. These insights are only visible when explanations are evaluated quantitatively.

Overall, this study establishes a clear and reproducible methodology for benchmarking audio \gls{xai} methods and provides the first evidence-driven assessment of their reliability in \gls{asd}. By moving beyond subjective visual interpretation and grounding explanation quality in behavioural validation, the proposed framework lays a foundation for more trustworthy, interpretable, and diagnostically meaningful machine-sound analysis systems.

\section*{Acknowledgment}

The authors would like to acknowledge the support of the Royal Air Force in the sponsoring of this research in support of the UK National Digital Twin Programme.

\bibliographystyle{IEEEtran}
\bibliography{refs}  






\end{document}